\documentclass[a4paper,11pt,aps,pra,superscriptaddress]{revtex4}
\def\be{\begin{equation}}
\def\ee{\end{equation}}
\def\beq{\begin{equation}}
\def\eeq{\end{equation}}
\def\bea{\begin{eqnarray}}
\def\eea{\end{eqnarray}}
\def\bf{\begin{figure}}
\def\ef{\end{figure}}
\def\bc{\begin{center}}
\def\ec{\end{center}}
\def\no{\nonumber}

\usepackage{graphicx}

\usepackage[T1]{fontenc}
\usepackage[bitstream-charter]{mathdesign} 

\begin{document}

\title{Aspects of coherent states of nonlinear algebras}
\author{T. Shreecharan}
\email{shreet@cts.iisc.res.in}
\affiliation{Centre for High Energy Physics, Indian Institute of Science, Bangalore, India 560 012}
\affiliation{The Institute of Mathematical Sciences, C.I.T. Campus, Taramani, Chennai, India 600 113}
\author{K. V. S. Shiv Chaitanya}
\email{chaitanya@imsc.res.in}
\affiliation{The Institute of Mathematical Sciences, C.I.T. Campus, Taramani, Chennai, India 600 113}
\begin{abstract}
\begin{center}
{\small ABSTRACT}
\end{center}
%
Various aspects of coherent states of nonlinear $su(2)$ and $su(1,1)$ algebras are studied. It is shown that the nonlinear $su(1,1)$ Barut-Girardello and Perelomov coherent states are related by a Laplace transform. We then concentrate on the derivation and analysis of the statistical and geometrical properties of these states. The Berry's phase for the nonlinear coherent states is also derived.
%
\end{abstract}

\maketitle

\section{Introduction}

Nonlinear algebras are interesting generalisations of Lie algebras. They made their first appearance in the study of curved Kepler and oscillator problems \cite{Higgs,Leemon}. To be precise the specific nonlinearity that Higgs encountered was of the cubic kind. Hence cubic algebras are also known as the Higgs algebras (HA). If the deformation is of degree two then such algebras are called quadratic algebras and were first considered by Sklyanin \cite{Skl}. After their arrival on the physics scene, these algebras have been studied by many authors and have made their appearance in very diverse areas.

They have been used to extend the gauge symmetry and thereby giving rise to a generalisation of Yang-Mills gauge theories \cite{Schoutens}. The dynamical symmetry associated with the two body Calogero model has been shown to be a HA \cite{Floreanini}. Staying with the Calogero model it was shown that this system possesses a hidden supersymmetry that is nonlinear \cite{Plyushchay}. Here it must be pointed out that Plyushay and co-workers have extensively studied this quantum mechanical nonlinear supersymmetry and have extended it to various other systems of physical interest \cite{Plyushchay2}. Further, nonlinear algebras have appeared in the study of identical particle symmetry in two dimensions \cite{Myrheim}, super-integrable systems in two dimensions \cite{Bonatsos1,Bonatsos2,Dask,Marquette}, multiphoton processes \cite{Debergh,Jagan}, quantum dot problems \cite{Gritsev} and several other places. Recently it has been proposed as a model for fuzzy space exhibiting topology change \cite{TRG}.

Apart from these physical applications, there have been studies from an abstract point of view \cite{Smith} and also from a mathematical physics perspective. The interest in these algebras is due the fact that one can construct finite as well as infinite dimensional (DIM) unitary irreducible representations (REPS). Finite DIM REPS occur when the algebra is a deformation of the $su(2)$ algebra and infinite DIM REPS occur when the $su(1,1)$ deformation is considered. Many interesting features of its representation theory have been studied by various authors \cite{Daskaloyannis,Bonatsos3,Beckers,Curtright,Rocek,Quesne1,Quesne2,Quesne3,Sunil1,Sunil2}.

In the present work, nonlinear algebras that are deformed versions of the angular momentum algebra and the simplest non-compact group, $SU(1,1)$ are considered. The various insights offered by the coherent states (CS) constructed for these deformations of the linear algebras are specifically studied. The construction of CS themselves is an involved thing and just a naive application of the prescription of Perelomov (P) \cite{Perelomov} or Barut-Girardello (BG) \cite{Barut} does not work here. The CS for the cubic angular momentum have been constructed in \cite{Jagan,Sadiq1} and for the cubic $su(1,1)$ it has been constructed in \cite{Sadiq2}. The CS for quadratically deformed $su(2)$ are worked out in \cite{Jagan,Sunilthese} and those of the $su(1,1)$ have been put forth in \cite{Sunilthese,Cannata}.

The presentation in the manuscript is as follows, the section II, the representation theory and the CS for the nonlinear $su(2)$ and $su(1,1)$ algebra is presented. Having provided the expressions of various types of CS in the previous section, we then show that the deformed $su(1,1)$ BGCS and PCS are related by a Laplace transform. In section IV statistical properties like, the photon number distribution, mean photon number, intensity correlation and the Mandel parameter, are calculated for each type of CS. For the special case of the cubic algebras and the linear algebras the plots for the same are given. The geometrical structure associated with these CS concerns section V. The Berry phase calculation for those Hamiltonians whose eigenstates states are given by the nonlinear $su(2)$ and $su(1,1)$ CS is the content of section V1. Results and discussion follow in section VI.

\section{Polynomial Algebras: Representation and CS} 

Nonlinear algebras can roughly be divided into two parts, one dealing with polynomial angular momentum algebra and the other polynomial $su(1,1)$ algebra. The former has finite DIM REPS and the latter infinite DIM REPS. Firstly, the polynomial $su(2)$ algebra is presented and then the polynomial $su(1,1)$ algebra is discussed.

\subsection{Polynomial $su(2)$ algebra and its CS}

Polynomial deformations of $su(2)$ algebra are characterised by the following commutation relations
\be \label{polsu2}
[J_+, J_-] = P(J_0) \equiv g(J_0) - g(J_0-1) \ , \qquad [J_0, J_\pm] = \pm \, J_\pm, 
\ee 
$g$'s are called the structure functions and their usefulness lies in the fact that they can be used to pin down the Casimir in an almost trivial manner
\be \label{polsu2cas}
\mathcal{J} = \frac{1}{2}\left[\{J_+, J_-\} + g(J_0) + g(J_0-1)\right].
\ee 
Similar to the $su(2)$ algebra, the finite DIM REPS are characterised by an integer or half integer $j$ of dimension $2j~+~1$. By considering a basis in which both the Casimir $\mathcal{J}$ and $J_0$ are diagonal
\be
\mathcal{J} \ \vert j, m  \rangle = g(j) \ \vert j, m  \rangle \qquad J_0 \ \vert j, m  \rangle = m \ \vert j, m  \rangle,
\ee
the action of the ladder operators is as follows
\be
J_+ \, \vert j, m  \rangle  =   \sqrt{g(j) - g(m)} \, \vert j, m+1  \rangle, \qquad
J_- \, \vert j, m  \rangle  =   \sqrt{g(j) - g(m-1)} \, \vert j, m+1  \rangle.
\ee
With an eye on later calculations, an arbitrary state $\vert j, m  \rangle$ is expressed  in terms of a Fock type basis. For this we choose $m=-j+n$, with this identification $\vert j, m  \rangle \equiv \vert j, -j + n  \rangle$. In the new basis the step operators act according to
\bea \label{su2ladfock}
J_0 \, \vert j, n  \rangle  & = &  (-j + n) \, \vert j, n  \rangle \\ \no
J_+ \, \vert j, n  \rangle  & = &   \sqrt{g(j) - g(-j+n)} \, \vert j, n+1  \rangle, \\ \no
J_- \, \vert j, n  \rangle  & = &   \sqrt{g(j) - g(-j+n-1)} \, \vert j, n-1  \rangle.
\eea
In the above equations and in what follows we supress $-j$ dependence in the second entry of the ket vectors. Let us mention here that new finite DIM REPS other than those presented in equation (\ref{su2ladfock}) have been constructed in \cite{Beckers}. By defining
\be  \label{genfacsu2}
\psi_n = g(j) - g(-j+n-1) \quad \mathrm{and} \quad [\psi_n]! = \prod_{\ell =1}^{n} \psi_\ell,
\ee
the ladder operators in (\ref{su2ladfock}) can be written in a more compact form
\be
J_+ \, \vert j, n  \rangle = \sqrt{\psi_{n+1}} \, \vert j, n+1  \rangle \qquad J_- \, \vert j, n  \rangle = \sqrt{\psi_{n}} \, \vert j, n-1  \rangle.
\ee
Thus, a general state $\vert j, n  \rangle$ can be constructed from the ground state in the following manner
\be
J_+^n \vert j, 0 \rangle = \sqrt{[\psi_n]!} \ \vert j, n  \rangle.
\ee
Note that $[\psi_0]!=1$. It must be pointed out that the discussion so far is valid for any polynomial deformation of the angular momentum algebra.

In what follows, we will be concerned only with the deformations that are odd degree $2p-1 (p=1,2, \cdots)$ for the polynomial;
\be \label{polsu2exp}
P(J_0) = 2 \sum_{r=1}^p \alpha_r J_0^r \sum_{s=1}^r (J_0+1)^{r-s} \ (J_0 - 1)^{s-1}.
\ee
$\alpha_r$'s are some real non-zero parameters. For a such a polynomial, the structure function is not difficult to find and is
\be
g(J_0) = \sum_{r=1}^p \alpha_r [J_0 (j_0+1)]^r.
\ee
Using the above in equation (\ref{genfacsu2}) we get
\be
\psi_n = \sum_{r=1}^p \alpha_r \ [j^r(j+1)^r + (j-n+1)^r (j-n)^r ],
\ee
which can be further cast into
\be \label{genfac2}
\psi_n = n \ (2j+1-n) \ \chi_n, \qquad [\psi_n]! = n! \ (2j+1-n)! \ [\chi_n]! \ ,
\ee
where
\be \label{dffacsu2}
\chi_n = \sum_{r=1}^p \sum_{s=1}^r \alpha_r \ [j(j+1)]^{r-s} \ [(j - n)(j - n + 1)]^{s-1}.
\ee
Casting the structure function in the form as given in Eq. (\ref{genfac2}), has the advantage that the contribution due to the nonlinear terms of the algebra is completely encoded in $\chi_n$, which in the linear limit goes to one. The above can be written in a more illuminating manner by factorising it in $n$, which is a polynomial of degree $2p-2$
\be
\chi_n = \alpha_p \ (n-a_1) \ (n-a_2) \cdots (n-a_{2p-2}), \qquad [\chi_n]! = \alpha_p^n \prod_{i=1}^{2p-2} (1-a_i)_n.
\ee
In the above $(a)_n$ is the Pochhammer symbol and is defined by
\be
(a)_n = a \ (a+1) \ (a+2) \cdots (a+n-1) = \frac{\Gamma(a+n)}{\Gamma(a)}, \qquad (a)_0 = 1.
\ee

We are now in a position to present the CS. Since, one is dealing with finite DIM REPS, one cannot construct an annihilation operator eigenstate. Another construction, that satisfies the resolution of identity as well as the overcompleteness property, for compact groups, was put forward by Perelomov \cite{Perelomov} and Gilmore \cite{Gilmore}. These type of states are known as PCS or the displacement operator states. It must be mentioned that a naive application of the displacement operator will not work in the present case since the algebra is nonlinear. One then has to modify the Perelomov technique suitably to get the right CS \cite{Jagan,Sunil2,Sadiq1}.

For the odd polynomial $su(2)$ algebra, given by Eq. (\ref{polsu2exp}), the PCS is given as \cite{Sadiq1}
\be \label{pcssu2}
\vert j,\zeta \rangle = N_p^{-\frac{1}{2}}(x) \sum_{n=0}^{2j}
\left(\begin{array}{c} 
2j \\ n
\end{array}\right)^\frac{1}{2}\sqrt{[\chi_n]!} \ \zeta^n \ \vert j, n\rangle
\ee
The normalisation factor is
\be \label{normnlinsu2}
|N_p(x)| = {_{2p-1} F_{0}[-2j, 1-a_1, 1-a_2, \cdots, 1-a_{2p-2};-;-x]}.
\ee
Here $x=\alpha_p|\zeta|^2$ and ${{_p} F_q}$ are the generalised hypergeometric series (GHS).

Let us set $p=2$ in Eq. (\ref{polsu2exp}) so that the polynomial is cubic or the Higgs algebra
\be
P(J_0) = 2 J_0 + 4 \alpha_2 J_0^3,
\ee
and the deformation factor of Eq. (\ref{dffacsu2}) turns out to be
\be \label{dffacHsu2}
\chi_n = 1 + \alpha_2 \ \Big[n^2 - (2j+1) n + 2j(j+1) \Big].
\ee
Note that in the above two equations we have set $\alpha_1=1$. Hermiticity requirement of the step operators yields $\alpha_2 \geq - 1/2j^2$. Factorising the deformation actor, Eq. (\ref{dffacHsu2})
\be
\chi_n = \alpha_2 \ (n - a_+) \ (n-a_-),
\ee
with the roots
\be \label{Hrootssu2}
a_\pm = \frac{1}{2} \left[(2j+1) \pm \sqrt{(2j+1)^2 - 8 j(j+1) - 4/\alpha_2 } \right].
\ee
We choose $\alpha_2 = 2$ which when substituted in Eq. (\ref{Hrootssu2}) leads to
\be \label{FHrootssu2}
a_\pm = \frac{1}{2} (2j+1) \ (1 \pm i).
\ee
The normalisation constant then becomes 
\be \label{FHnormnlinsu2}
|N_2(x)| = {_3F_{0}[-2j, 1-a_+, 1-a_-;-;-x]}.
\ee
This completes the finite dimensional case.

\subsection{Polynomial $su(1,1)$ algebra and its CS}

Polynomial deformations of $su(1,1)$ algebra  are characterised by the following commutation relations
\be \label{polsu11}
[K_+, K_-] = P(K_0) \equiv g(K_0-1) - g(K_0) \ , \quad [K_0, K_\pm] = \pm \, K_\pm. 
\ee 
The Casimir for the deformed $su(1,1)$ algebra acquires the form
\be \label{polsu11cas}
\mathcal{K} = \frac{1}{2}\left[g(K_0) + g(K_0-1)- \{K_+, K_-\} \right].
\ee 
The REPS are infinite dimensional just like the usual $su(1,1)$ algebra. Hence, a basis $\{\vert k, p  \rangle\}$ can be constructed such that it is a simultaneous eigenstate of $\mathcal{K}$ and $K_0$
\be
\mathcal{K} \, \vert k, p  \rangle  =  g(k-1) \, \vert k, p  \rangle \qquad K_0 \, \vert k, p  \rangle  =  p \, \vert k, p  \rangle.
\ee
The raising and lowering operators act according to 
\be
K_+ \, \vert k, p  \rangle  =   \sqrt{g(p) - g(k-1)} \, \vert k, p+1  \rangle, \qquad
K_- \, \vert k, p  \rangle  =   \sqrt{g(p-1) - g(k-1)} \, \vert k, p-1  \rangle.
\ee
Similar to the deformed $su(2)$ case, $su(1,1)$ states are expressed in terms of an integer basis. This can be achieved via the identification $p = k+n$. The action of the generators on the new basis is
\bea \label{gcd}
K_0 \, \vert k, n  \rangle  & = &  (k + n) \, \vert k, n  \rangle \\ \no
K_+ \, \vert k, n  \rangle  & = &   \sqrt{g(k + n) - g(k-1)} \, \vert k, n+1  \rangle, \\ \no
K_- \, \vert k, n  \rangle  & = &   \sqrt{g(k + n -1) - g(k-1)} \, \vert k, n-1  \rangle.
\eea
In the above equations and in what follows we will supress $k$ in the second entry of the nonlinear $su(1,1)$ ket vectors. Let us define
\be \label{phin}
\phi_n = g(k+n-1) - g(k-1) \quad \mathrm{and} \quad [\phi_n]! = \prod_{\ell =1}^{n} \phi_\ell(k) \ .
\ee 
With this change Eq. (\ref{gcd}) can be written in the compact form
\be
K_+ \, \vert k, n  \rangle = \sqrt{\phi_{n+1}} \, \vert k, n+1  \rangle \quad K_- \, \vert k, n  \rangle = \sqrt{\phi_{n}} \, \vert k, n-1  \rangle.
\ee
An arbitrary state $\vert k, n  \rangle$ can be constructed by the repeated application of the raising operator on the ground state
\be \label{r1}
K_+^n \ \vert k,0 \rangle = \sqrt{[\phi_n]!} \ \vert k, n \rangle.
\ee
So far what has been discussed is independent of the degree of the polynomial. In what follows we will specialise to the case of the odd deformations of $su(1,1)$ and present its CS.

A general polynomial of odd degree $2p-1$, for $p=1, 2, \cdots$ is 
\be \label{polsu11exp}
P(K_0) = -2 \sum_{r=1}^p \beta_r K_0^r \sum_{s=1}^r (K_0 + 1)^{r-s} \ (K_0 - 1)^{s-1}.
\ee
$\beta_r$'s are non-zero real parameters. For this polynomial deformation, just like the $su(2)$ case, the structure function is found to be
\be
g(K_0) = \sum_{r=1}^p \beta_r [K_0 (K_0+1)]^r.
\ee
Using the above in Eq. (\ref{phin}) we obtain
\be
\phi_n = \sum_{r=1}^p \beta_r \ \Big[(k+n)^r (k+n-1)^r - k^r (k-1)^r \Big].
\ee
Again, just like previous sub-section we can write the above in a manner that clearly brings out the nonlinear contributions:
\be
\phi_n = n \ (2k - 1 + n) \ \rho_n \ , \qquad [\phi_n]! = n! \ (2k)_n \ [\rho_n]! \ ,
\ee
where
\be \label{dffacsu11}
\rho_n = \sum_{r=1}^p \sum_{s=1}^r \beta_r \ [k(k - 1)]^{r-s} \ [(k + n)(k + n - 1)]^{s-1},
\ee
or
\be
\rho_n = \beta_p \ (n-b_1) \ (n-b_2) \cdots (n-b_{2p-2}), \qquad [\rho_n]! = \beta_p \prod^{2p-2}_{i=1} (1-b_i)_n.
\ee
$b$'s are roots of the equation $\rho_n = 0$.

In the present case, unlike the deformed su(2), one can construct two different types of CS. One is the PCS and the other an eigenstate of the lowering operator. This latter CS also goes by the name of Barut-Girardello (BG) \cite{Barut}. We will not provide the explicit construction of these states and the reader is referred to \cite{Sunil2,Jagan,Sadiq2} for details. 

The nonlinear BGCS is \cite{Sadiq2}
\be \label{bgcs}
\vert k, \xi  \rangle = N_p^{-\frac{1}{2}}(y) \sum_{n=0}^{\infty} \frac{(\xi)^n}{\sqrt{[\phi_n]!}} \ \vert k,n . \rangle 
\ee
and the normalisation factor is found to be
\be \label{nlinbgcsnorm}
|N_p(y)| = {_0F_{2p-1}[-;2k,1-b_1,1-b_2, \cdots, 1-b_{2p-1};y]},
\ee
with $y = |\xi|^2 / \beta_p$. The PCS is of the form \cite{Sadiq2}
\be \label{pcssu11}
\vert k,\eta \rangle = N_p^{-\frac{1}{2}}(z) \sum_{n=0}^\infty \left[\sqrt{\frac{(2k)_n}{ n! [\rho_n]!}} \ \right] \ \eta^n \, \vert k,n \rangle.
\ee
The normalisation is 
\be \label{nlinpcsnorm}
|N_p(z)| = {_1F_{2p-2} [2k; 1-b_1,1-b_2, \cdots, 1-b_{2p-2};z]},
\ee
with $z = |\eta|^2/\beta_p $.

Similar to the $su(2)$ case, set $p=2$ in Eq. (\ref{polsu11exp}) so that the polynomial is cubic
\be
P(K_0) = - 2 K_0 - 4 \beta_2 K_0^3,
\ee
and the deformation factor of Eq. (\ref{dffacsu11}) takes the form
\be \label{dffacHsu11}
\rho_n = 1 + \beta_2 \ \Big[n^2 + (2k-1) n + 2k(k-1) \Big].
\ee
In the above equations we have set $\beta_1=1$. Eq. (\ref{dffacHsu11}) can be factorised
\be
\rho_n = \beta_2 \ (n - b_+) \ (n - b_-),
\ee
with the roots being
\be \label{Hrootssu11}
b_\pm = - \frac{1}{2} \left[(2k-1) \pm \sqrt{(2k-1)^2 - 8 k(k-1) - 4/\beta_2 } \right].
\ee
As before we choose $\beta_2 = 2$ which when substituted in Eq. (\ref{Hrootssu11}) leads to
\be \label{FHrootssu11}
b_\pm = - \frac{1}{2} (2k-1) \ (1 \pm i).
\ee
The normalisation constant for the cubic $su(1,1)$ BGCS then becomes
\be \label{FHnormnlinsu11BGCS}
|N_2(y)| = {_0F_{3}[-;2k, 1-b_+, 1-b_-;y]}.
\ee
The corresponding normalisation for the PCS is
\be \label{FHnormnlinsu11PCS}
|N_2(z)| = {_1F_{2}[2k; 1-b_+, 1-b_-;z]}.
\ee

\section{Laplace transform between the PCS and BGCS}

In this section we show that the PCS and the BGCS, presented in the previous section, for the nonlinear $su(1,1)$ algebra are related via a Laplace transform. We begin by taking a normalized state
\be \label{nor}
\vert \psi  \rangle = \sum_{n=0}^{\infty}c_n  \vert k,n  \rangle,
\ee
and construct an analytic function $F(\xi,k)$ with the BGCS (\ref{bgcs}) as
\be \label{l1}
F(\xi,k) = \sum_{n=0}^{\infty}c_n \left(\frac{\Gamma(2k)}{n!\Gamma(n+2k)[ \rho_n]!}\right)^\frac{1}{2}(\xi)^n.
\ee
Here
\be
(2k)_n=\frac{\Gamma(n+2k)}{\Gamma(2k)}.
\ee
Using equation (\ref{nor}) and PCS (\ref{pcssu11}), we define a new function $G(\eta,k)$ by
\be
G(\eta,k)= N_p^{1/2}(z)\langle \psi \vert  \eta,k  \rangle = \sum_{n=0}^{\infty}c_n \left(\frac{\Gamma(n+2k)}{n!\Gamma(2k)[ \rho_n]!}\right)^\frac{1}{2}(\eta)^n \ .
\ee
Set $\eta =1/Z$ so that
\be \label{l2}
G(\frac{1}{Z},k)= \sum_{n=0}^{\infty}c_n \left(\frac{\Gamma(n+2k)}{n!\Gamma(2k)[ \rho_n]!}\right)^\frac{1}{2}(\frac{1}{Z})^n \ .
\ee
The integral form for the gamma function is given by
\be \label{l3}
\Gamma(n+2k)=Z^{2k+n} \int_0^\infty d\xi  \ \xi^{2k+n-1} e^{- Z \xi} \ .
\ee
Putting equation (\ref{l3}) in (\ref{l2}) and interchanging the summation and integral we get
\be \label{l4}
G(\frac{1}{Z},k)= \int_0^\infty d\xi \ \frac{Z^{2k}}{\sqrt{\Gamma(2k)}}\xi^{2k-1}e^{- Z \xi}\sum_{n=0}^{\infty}c_n \left(\frac{1}{n!\Gamma(n+2k) \ .
[ \chi_n]!}\right)^\frac{1}{2}(\xi)^n 
\ee
One can recognize that the sum is nothing but that given in Eq. (\ref{l1}), hence we finally get
\be \label{l5}
G(\frac{1}{Z},k)= \int_0^\infty d\xi \ \frac{Z^{2k}}{\sqrt{\Gamma(2k)}} \ \xi^{2k-1}F(\xi,k) \ e^{-z\xi} \ .  
\ee
From (\ref{l5}) we see that the BGCS state and PCS are related by a Laplace transform. The same result for the linear $su(1,1)$ case was first derived in \cite{Brif-Lap}.

\section{Statistical Properties of the Nonlinear CS}

Statistical properties of CS have applications in quantum optics, quantum electronics and in phenomenological models explaining some observable phenomena. The CS constructed in the previous section have interesting quantum statistical properties analogous to the classical radiation field. These properties give a clear insight into the deviation of the behaviour of the CS: classical to nonclassical. The classical features are given by the Poisson distribution and non-classical behaviour manifests as deviations form this. The deviations from classical behaviour can be measured with the Mandel parameter ($\mathcal{Q}$) \cite{Mandel} which vanishes for the Poisson distribution. The parameter is positive for a super-Poissonian distribution and is negative for sub-Poissonian. Another useful quantity that can be used is the intensity correlation function ($\mathcal{I}$) which if greater than $1$ has photon-bunching effect and for values less than $1$ we get photon-antibunching hence is sub-Poissonian. The traditional notation for intensity correlation is $g^{(2)}$, but since we have already used $g$ to denote the structure functions of the algebra we use $\mathcal{I}$ for the former quantity.

Let us denote a generic CS, whose normalisation constant is GHS ${{_p} F_q [\bar{a}_1, \cdots, \bar{a}_p; \bar{b}_1, \cdots, \bar{b}_q; \bar{x}]} \equiv N(\bar{x})$, by $\vert p;q;\alpha \rangle$. Here $\bar{x}$ is the variable and can be $x,y,z$; $\alpha$ is a complex parameter appearing in the CS which may be $\zeta,\xi,\eta$ as the case maybe. Such states were constructed and studied in \cite{Appl} which in turn were based on the study initiated by the work \cite{Klauder}. 

The overlap of GHCS with the number operator eigenstate state $\vert n \rangle$ is $ \langle n \vert p;q;\alpha \rangle$. The photon number distribution is then given by
\be \label{photondist}
\mathcal{P} = \big\vert \langle n \vert p;q;\alpha \rangle \big\vert^2 \ .
\ee
The mean photon number is
\be \label{mphotonumb}
\mathcal{N} = \bar{x} \ \frac{N^\prime(\bar{x})}{N(\bar{x})} \ .
\ee
The intensity correlation and the Mandel parameter are
\be \label{intcorr}
\mathcal{I} = \frac{N^{\prime \prime}(\bar{x}) \ N(\bar{x})}{N^{\prime 2}(\bar{x})} \ ,
\ee
and
\be \label{mandel}
\mathcal{Q} = \bar{x} \ \left(\frac{N^{\prime \prime}(\bar{x})}{N^{\prime}(\bar{x})}-\frac{N^\prime(\bar{x})}{N(\bar{x})} \right) \ ,
\ee
respectively. Note that the primes denote derivative with respect to $\bar{x}$.

In the subsections below we will calculate and plot the photon distribution, mean photon number, intensity correlation, and the Mandel parameter. In the next section the metric associated with the nonlinear CS is presented. Analytic expressions are given for the arbitrary odd polynomial CS and plots are included for the cubic and the linear CS. In all the plots provided in the present work we used: large dashes (blue) for $j/k = 1/2$, dots (green) for $j/k = 1$, dots-dashes (red) for $j/k = 3$, and dashes (purple) $j/k = 8$. Furthermore, these statistical and metric expressions have been obtained for the cubic $su(2)$ from equations (\ref{FHrootssu2}) and (\ref{FHnormnlinsu2}), for cubic $su(1,1)$ BGCS and PCS from equations (\ref{FHrootssu11}), (\ref{FHnormnlinsu11BGCS}), and (\ref{FHnormnlinsu11PCS}).

\subsection{Deformed $su(2)$}

The photon number distribution for the PCS, Eq. (\ref{pcssu2}) is
\be
\mathcal{P}(x) = N_p^{-1}(x) \left(\begin{array}{c} 
2j \\ n
\end{array}\right) \prod_{i=1}^{2p-2} (1-a_i)_n \ x^n
\ee
the mean photon number is found to be
\be
\mathcal{N}(x) = -x \ (-2j) \ \frac{_{2p-1}F_{0}[-2j+1,2-a_1,2-a_2, \cdots, 2 - a_{2p-2};-;-x]}{_{2p-1}F_{0}[-2j,1-a_1,1-a_2, \cdots, 1-a_{2p-2};-;-x]} \ \prod_{i=1}^{2p-2} (1-a_i) \ .
\ee
The intensity correlation is given by
\bea \no
\mathcal{I}(x) = \frac{(-2j+1)}{(-2j)} \ \frac{\prod_{i=1}^{2p-2} (2-a_i)}{\prod_{i=1}^{2p-2} (1-a_i)} \ _{2p-1}F_{0}[-2j, 1-a_1, 1-a_2, \cdots, 1-a_{2p-2};-;-x] \\
\frac{_{2p-1}F_{0}[-2j+2, 3-a_1, 3-a_2, \cdots, 3-a_{2p-2};-;-x]}{_{2p-1}F_{0}[-2j+1, 2-a_1, 2-a_2, \cdots, 2-a_{2p-2};-;-x]^2} \ .
\eea
Using the expression of the normalisation constant of Eq. (\ref{normnlinsu2}) in equation (\ref{mandel}) leads to
\bea \no 
\mathcal{Q}(x) = -x \ \left[ (-2j+1) \ \frac{_{2p-1}F_{0}[-2j+2, 3-a_1, 3-a_2, \cdots, 3-a_{2p-2};-;-x]}{_{2p-1}F_{0}[-2j+1, 2-a_1, 2-a_2, \cdots, 2-a_{2p-2};-;-x]} \ \prod_{i=1}^{2p-2} (2-a_i) \right. \\ \left.
- (-2j) \ \frac{_{2p-1}F_{0}[-2j+1,2-a_1,2-a_2, \cdots, 2 - a_{2p-2};-;-x]}{_{2p-1}F_{0}[-2j,1-a_1,1-a_2, \cdots, 1-a_{2p-2};-;-x]} \ \prod_{i=1}^{2p-2} (1-a_i) \right] \ .
\eea
\bf
\bc
\begin{tabular}{cc}
\begin{minipage}{3in}
\centering
\includegraphics[height=2in,width=2.8in]{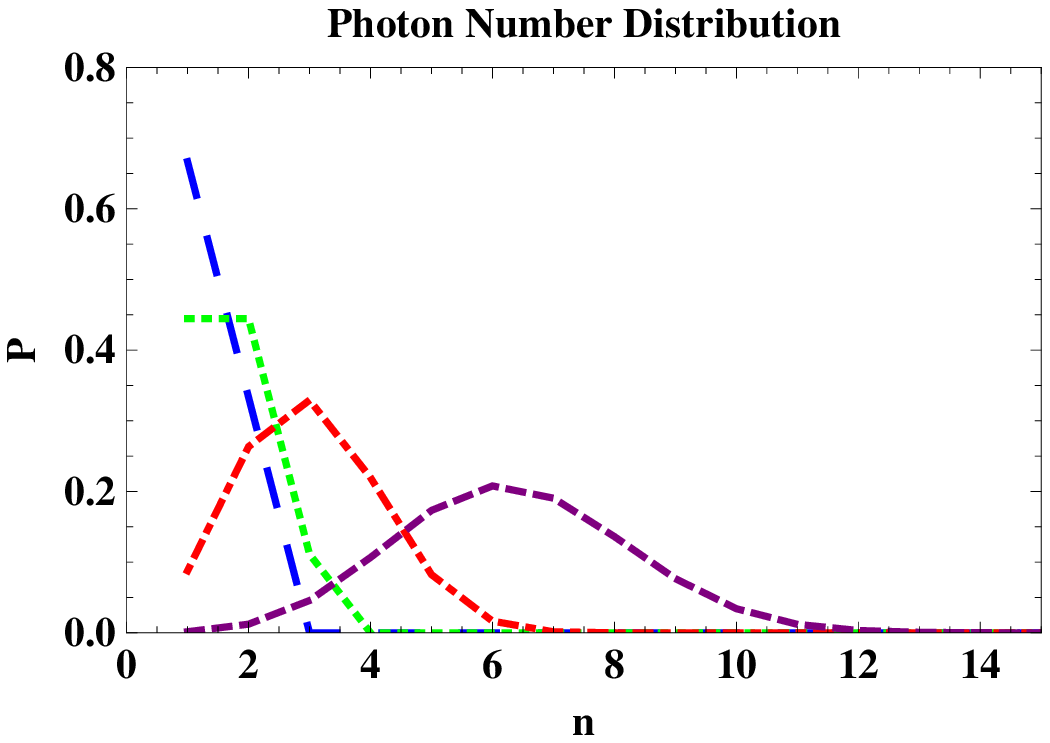}
\caption{$su(2)$ PCS} \label{su2photdist}
\end{minipage}
&
\begin{minipage}{3in}
\centering
\includegraphics[height=2in,width=2.8in]{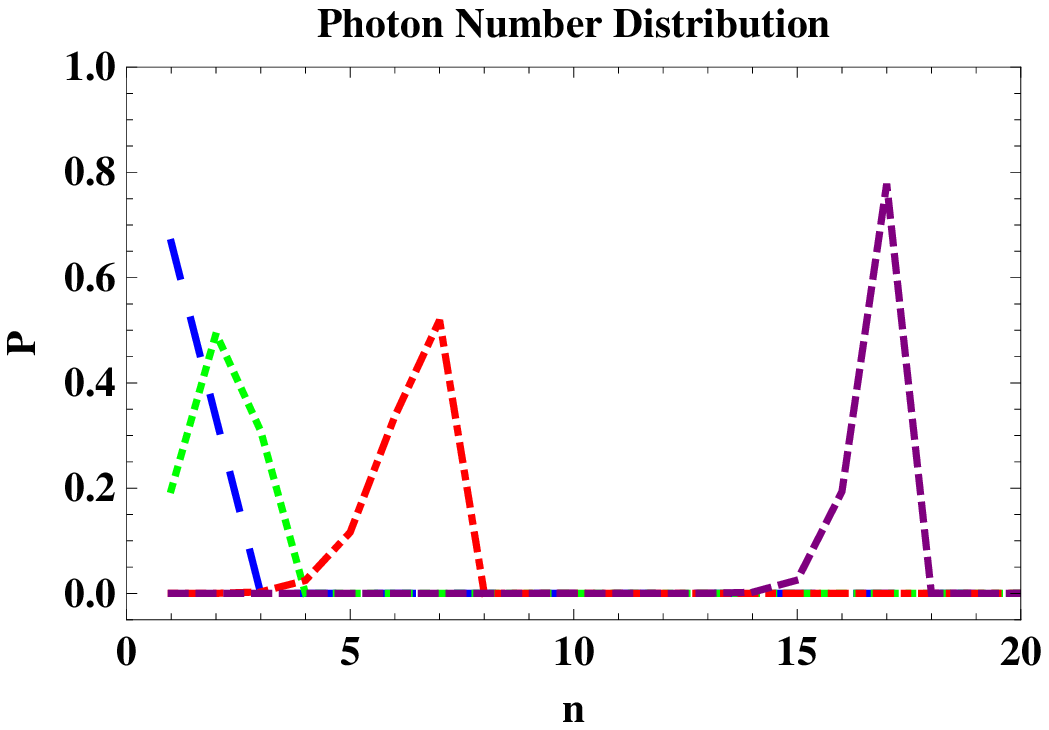}
\caption{Nonlinear $su(2)$ PCS} \label{Nsu2photdist}
\end{minipage}
\end{tabular}
\ec
\ef
\bf
\bc
\begin{tabular}{cc}
\begin{minipage}{3in}
\centering
\includegraphics[height=2in,width=2.8in]{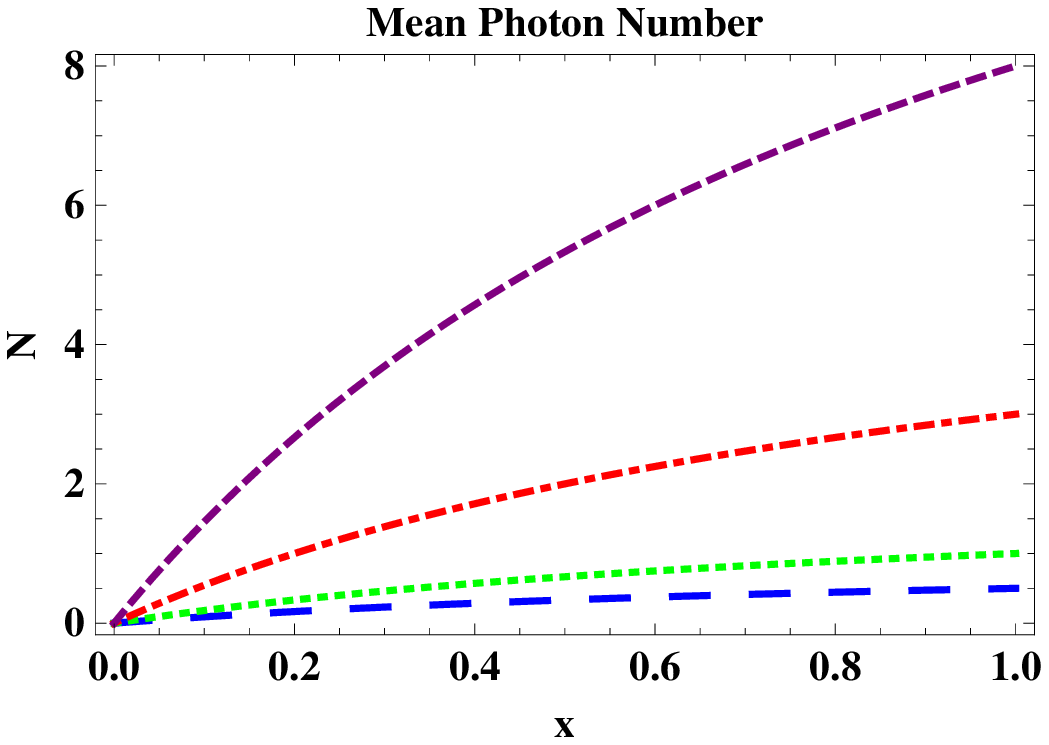}
\caption{$su(2)$ PCS} \label{su2mphotnumb}
\end{minipage}
&
\begin{minipage}{3in}
\centering
\includegraphics[height=2in,width=2.8in]{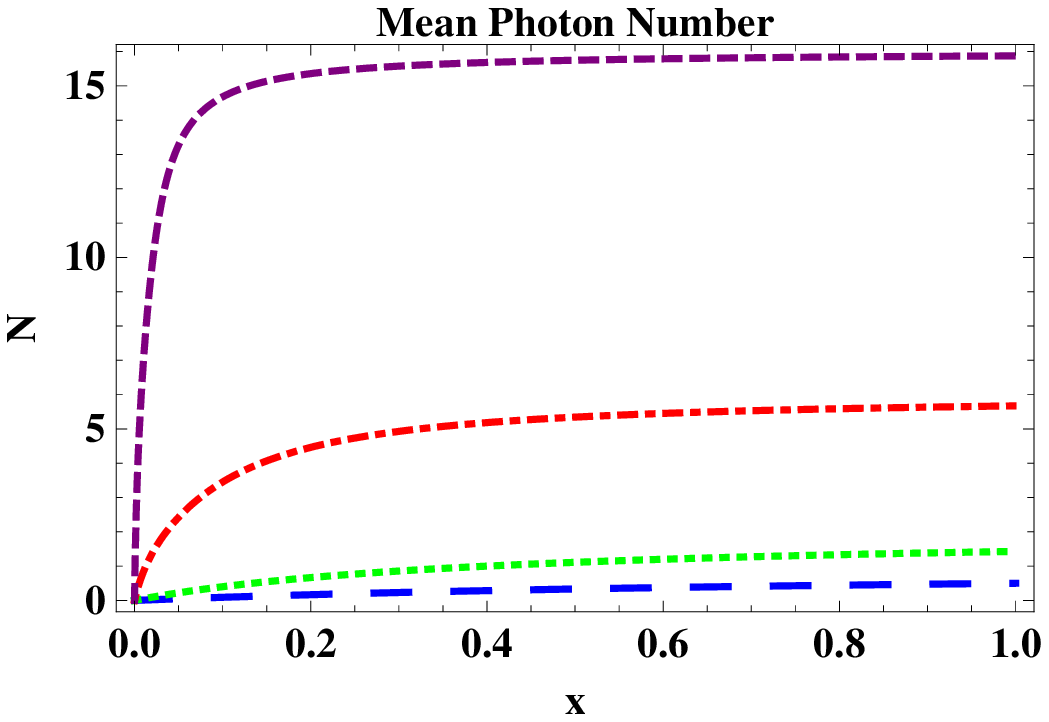}
\caption{Nonlinear $su(2)$ PCS} \label{Nsu2mphotnumb}
\end{minipage}
\end{tabular}
\ec
\ef
\bf
\bc
\begin{tabular}{cc}
\begin{minipage}{3in}
\centering
\includegraphics[height=2in,width=2.8in]{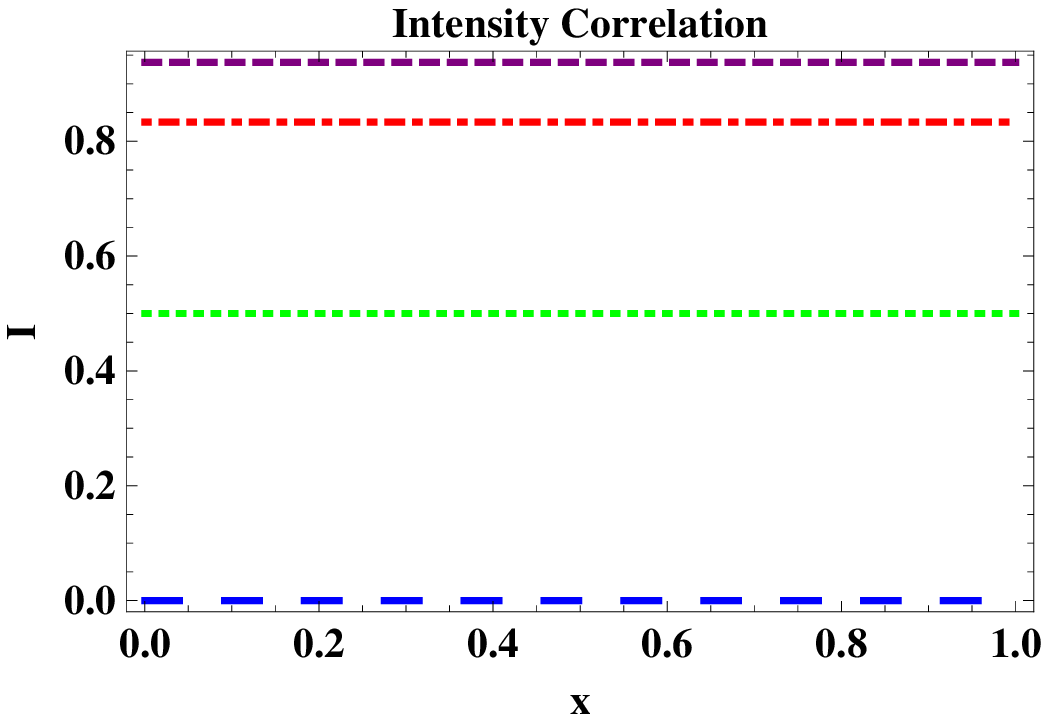}
\caption{$su(2)$ PCS} \label{su2intcorr}
\end{minipage}
&
\begin{minipage}{3in}
\centering
\includegraphics[height=2in,width=2.8in]{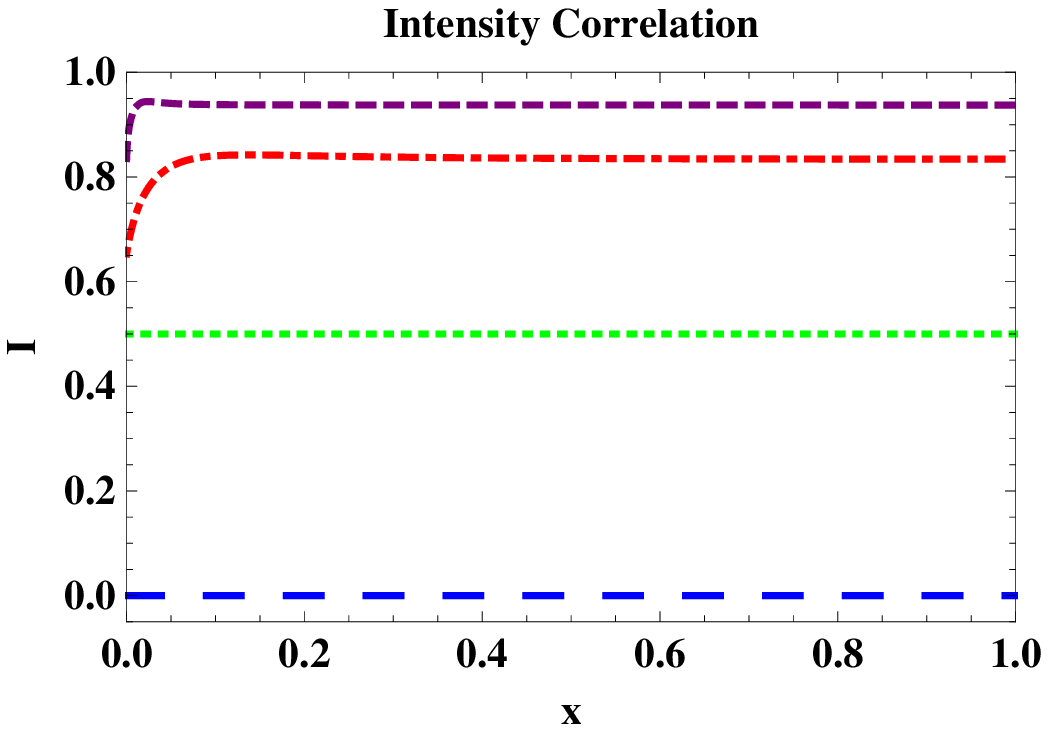}
\caption{Nonlinear $su(2)$ PCS} \label{Nsu2intcorr}
\end{minipage}
\end{tabular}
\ec
\ef
\bf
\bc
\begin{tabular}{cc}
\begin{minipage}{3in}
\centering
\includegraphics[height=2in,width=2.8in]{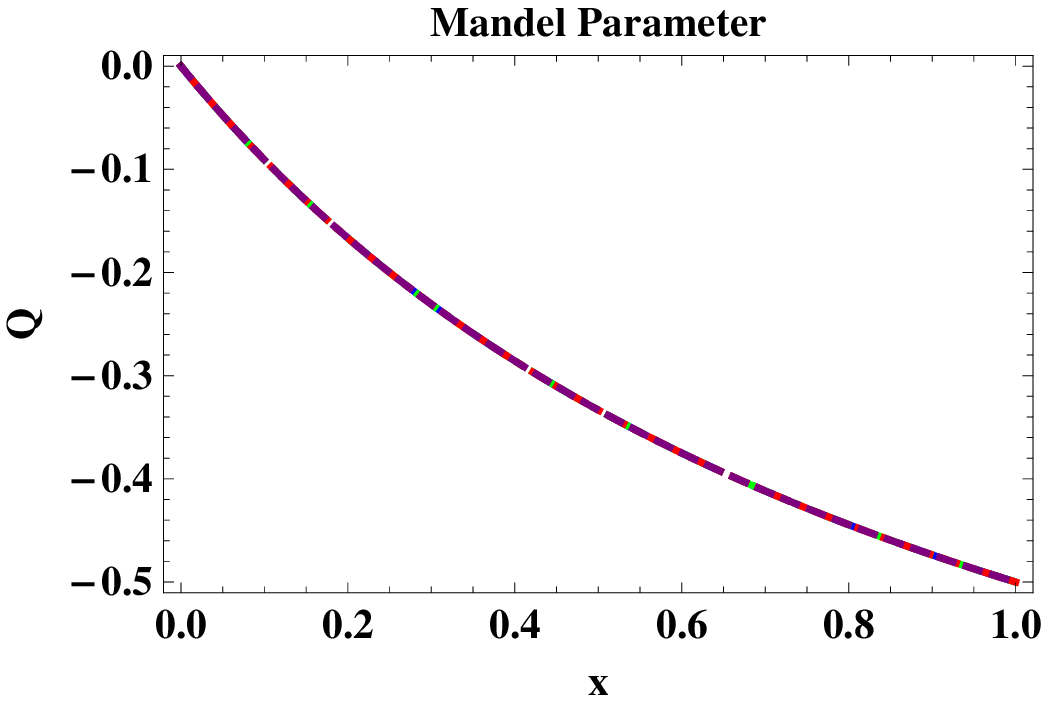}
\caption{$su(2)$ PCS} \label{su2mandel}
\end{minipage}
&
\begin{minipage}{3in}
\centering
\includegraphics[height=2in,width=2.8in]{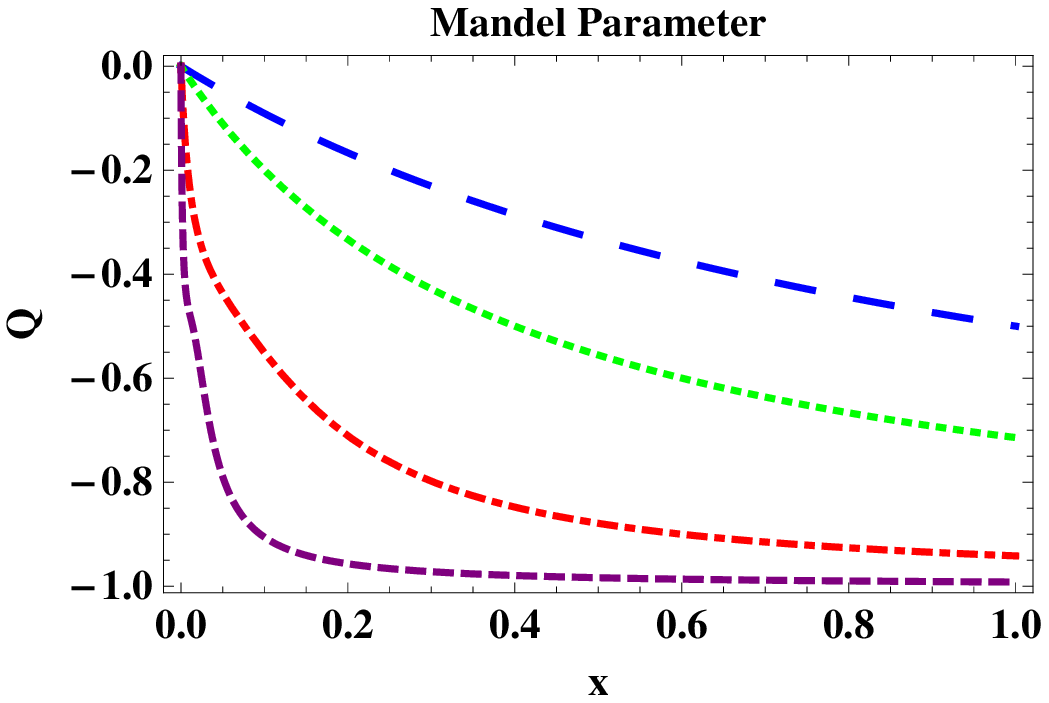}
\caption{Nonlinear $su(2)$ PCS} \label{Nsu2mandel}
\end{minipage}
\end{tabular}
\ec
\ef

The graphs of figures (\ref{su2photdist}), (\ref{su2mphotnumb}), (\ref{su2intcorr}), and (\ref{su2mandel}) are obtained from equations (\ref{photondist}), (\ref{mphotonumb}), (\ref{intcorr}) and (\ref{mandel}) respectively. These plots have been obtained using the  normalisation constant of $su(2)$ PCS which is $N_1(x) = {_1}F_0[-2j;-;-x]$. Figures (\ref{Nsu2photdist}), (\ref{Nsu2mphotnumb}), (\ref{Nsu2intcorr}), and (\ref{Nsu2mandel}) are the plots obtained from the same aforementioned equations for the corresponding nonlinear case: Eq. (\ref{FHnormnlinsu2}). It can be inferred that the statistics is sub-Poissonian, for both $su(2)$ and nonlinear $su(2)$, because form the intensity correlation plots Fig. (\ref{su2intcorr}) and Fig. (\ref{Nsu2intcorr}) it is clear that $\mathcal{I} < 1$. Also the same conclusion can be reached by seeing the Mandel parameter plots Fig. (\ref{mandel}) and Fig. (\ref{Nsu2mandel}): $\mathcal{Q} < 0$. It can be noticed that the Mandel parameter corresponding to $su(2)$ CS is independent of the $j$ value and for the nonlinear $su(2)$ CS it is dependent. Furthermore it can also be seen that the value approaches the value $-1$ for large values of $j$, signifying considerable deviation from Poisson behaviour thereby indicating strong non-classicality. The mean photon number is shown in figures (\ref{su2mphotnumb}) and (\ref{Nsu2mphotnumb}). For small $j$ both the linear and nonlinear results are similar but there is a marked difference as the $j$ value increases. In the case of photon number distribution plots Figs. (\ref{su2photdist}) and (\ref{Nsu2photdist}), the linear $su(2)$ CS behaviour is binomial but for the nonlinear case it is not.

\subsection{Deformed $su(1,1)$ BGCS}

The photon number distribution for Eq. (\ref{bgcs}) is
\be
\mathcal{P}(y) = \frac{N_p^{-1}(y)}{n! \ (2k)_n \prod_{i=1}^{2p-2} (1-b_i)_n} \ y^n \,
\ee
and the mean photon number is found to be
\be
\mathcal{N}(y) = \frac{y}{(2k)} \ \frac{_0F_{2p-1}[-; 2k+1, 2-b_1, 2-b_2, \cdots, 2-b_{2p-2};y]}{_0F_{2p-1}[-; 2k, 1-b_1, 1-b_2, \cdots, 1-b_{2p-2};y]} \ \frac{1}{\prod_{i=1}^{2p-2} (1-b_i)} \ .
\ee
Intensity correlation for the nonlinear BGCS turns out to be
\bea \no
\mathcal{I}(y) =   \frac{(2k)}{(2k+1)} \ \frac{\prod_{i=1}^{2p-2} (1-b_i)}{\prod_{i=1}^{2p-2} (2-b_i)} \ _0F_{2p-1}[-; 2k, 1-b_1, 1-b_2, \cdots, 1-b_{2p-2};y] \\ \frac{_0F_{2p-1}[-; 2k+2, 3-b_1, 3-b_2, \cdots, 3-b_{2p-2};y]}{_0F_{2p-1}[-; 2k+1, 2-b_1, 2-b_2, \cdots, 2-b_{2p-2};y]^2} \ .
\eea
The Mandel parameter 
\bea \no 
\mathcal{Q}(y) = y \ \left[\frac{1}{(2k+1)} \ \frac{_0F_{2p-1}[-; 2k+2, 3-b_1, 3-b_2, \cdots, 3-b_{2p-2};y]}{_0F_{2p-1}[-; 2k+1, 2-b_1, 2-b_2, \cdots, 2-b_{2p-2};y]} \ \frac{1}{\prod_{i=1}^{2p-2} (2-b_i)} \right. \\ \left. - \ \frac{1}{(2k)} \ \frac{_0F_{2p-1}[-; 2k+1, 2-b_1, 2-b_2, \cdots, 2-b_{2p-2};y]}{_0F_{2p-1}[-; 2k, 1-b_1, 1-b_2, \cdots, 1-b_{2p-2};y]} \ \frac{1}{\prod_{i=1}^{2p-2} (1-b_i)} \right] \ .
\eea
\bf
\bc
\begin{tabular}{cc}
\begin{minipage}{3in}
\centering
\includegraphics[height=2in,width=2.8in]{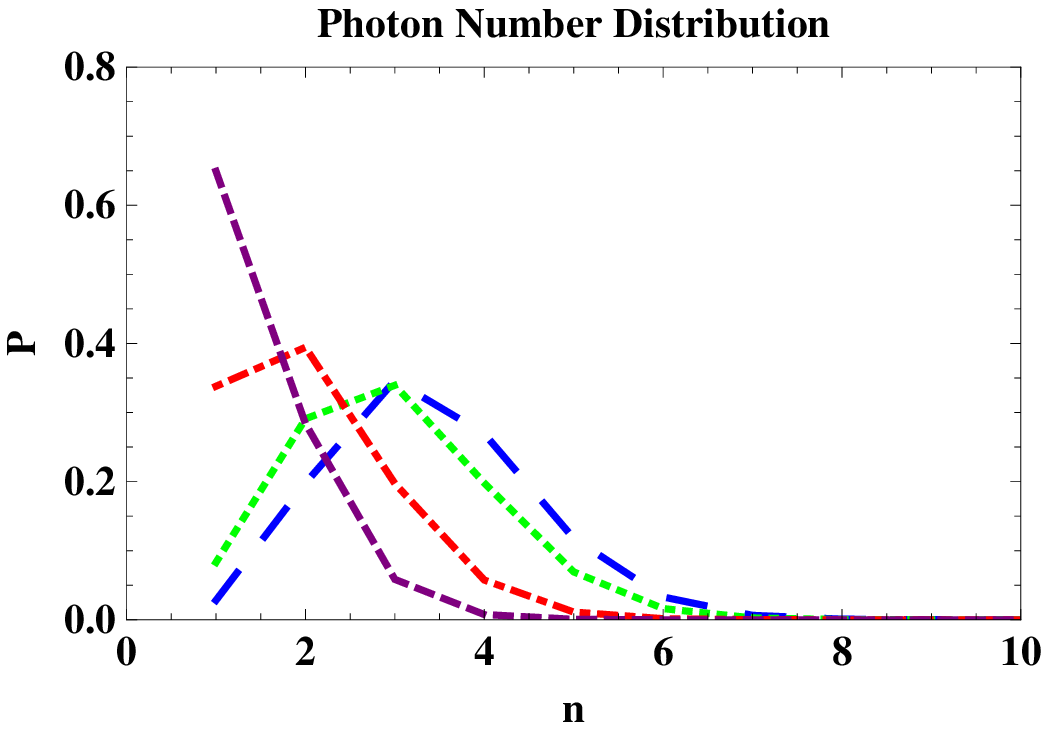}
\caption{$su(1,1)$ BGCS} \label{su11BGphotdist}
\end{minipage}
&
\begin{minipage}{3in}
\centering
\includegraphics[height=2in,width=2.8in]{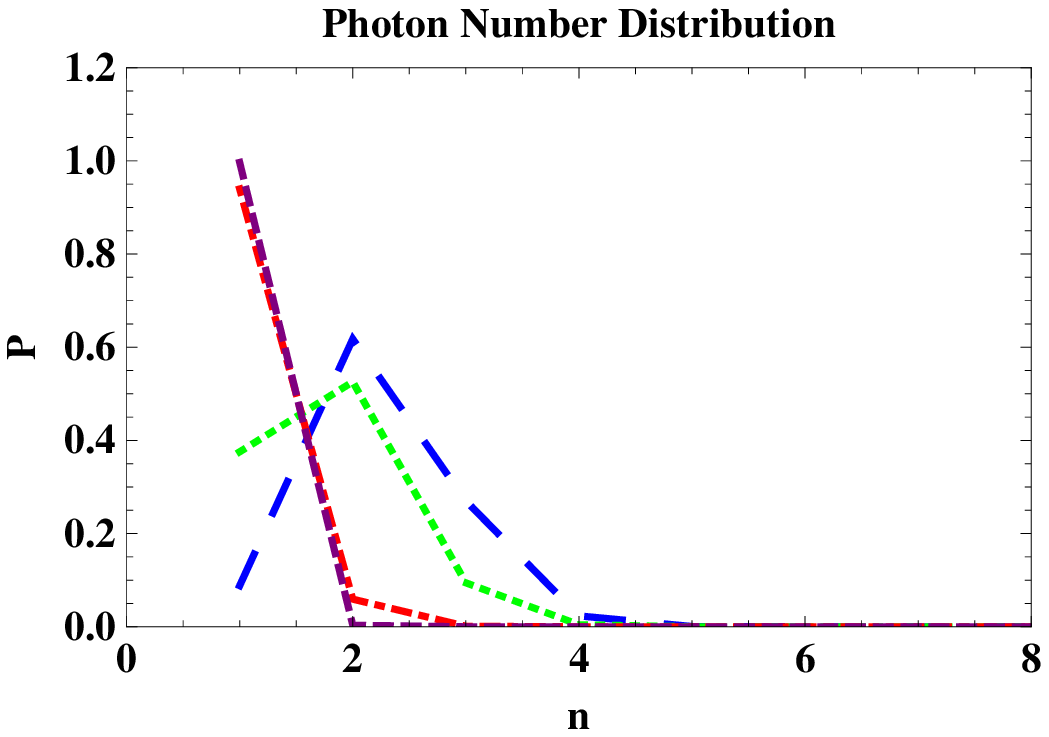}
\caption{Nonlinear $su(1,1)$ BGCS} \label{Nsu11BGphotdist}
\end{minipage}
\end{tabular}
\ec
\ef
\bf
\bc
\begin{tabular}{cc}
\begin{minipage}{3in}
\centering
\includegraphics[height=2in,width=2.8in]{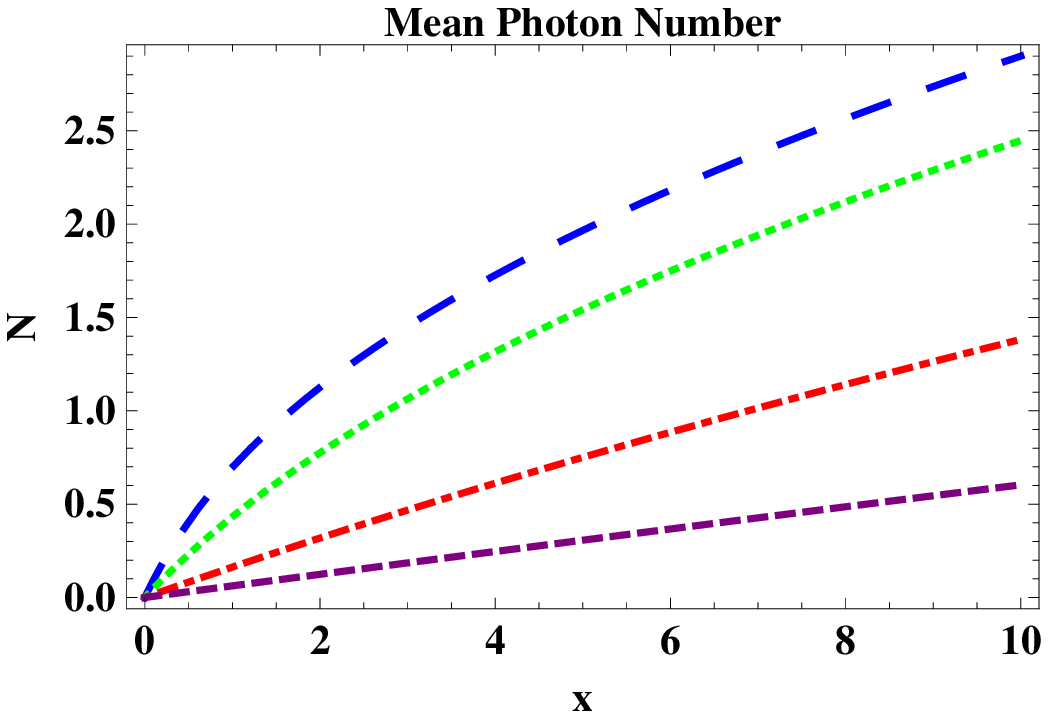}
\caption{$su(1,1)$ BGCS} \label{su11BGmphotnumb}
\end{minipage}
&
\begin{minipage}{3in}
\centering
\includegraphics[height=2in,width=2.8in]{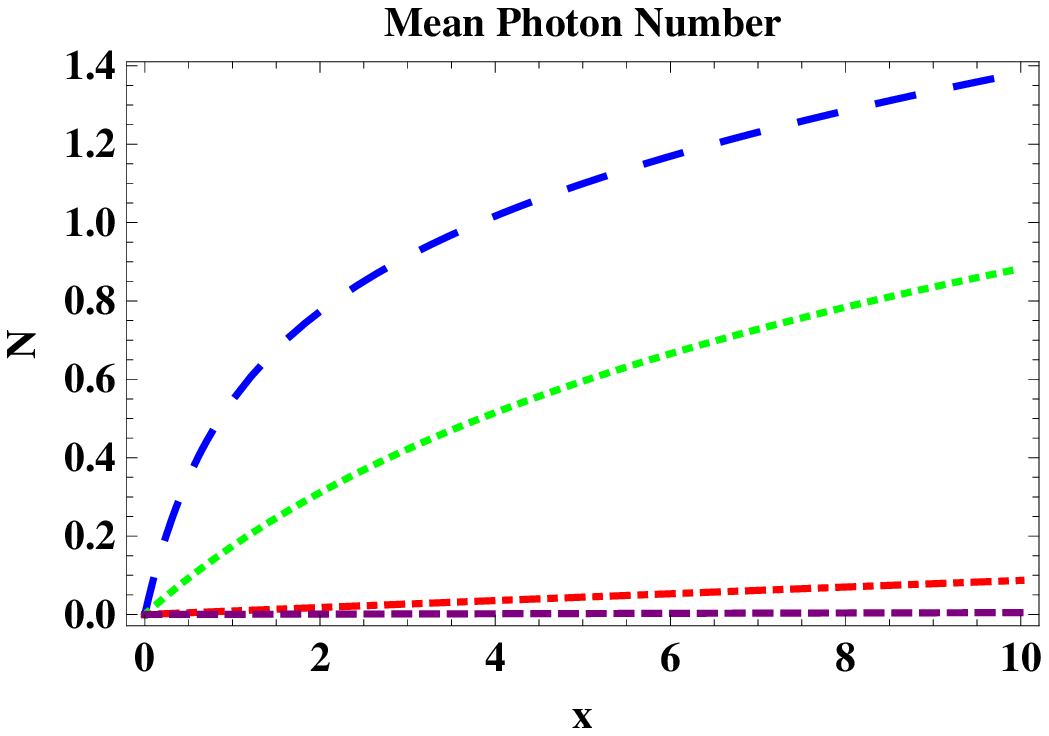}
\caption{Nonlinear $su(1,1)$ BGCS} \label{Nsu11BGmphotnumb}
\end{minipage}
\end{tabular}
\ec
\ef
\bf
\bc
\begin{tabular}{cc}
\begin{minipage}{3in}
\centering
\includegraphics[height=2in,width=2.8in]{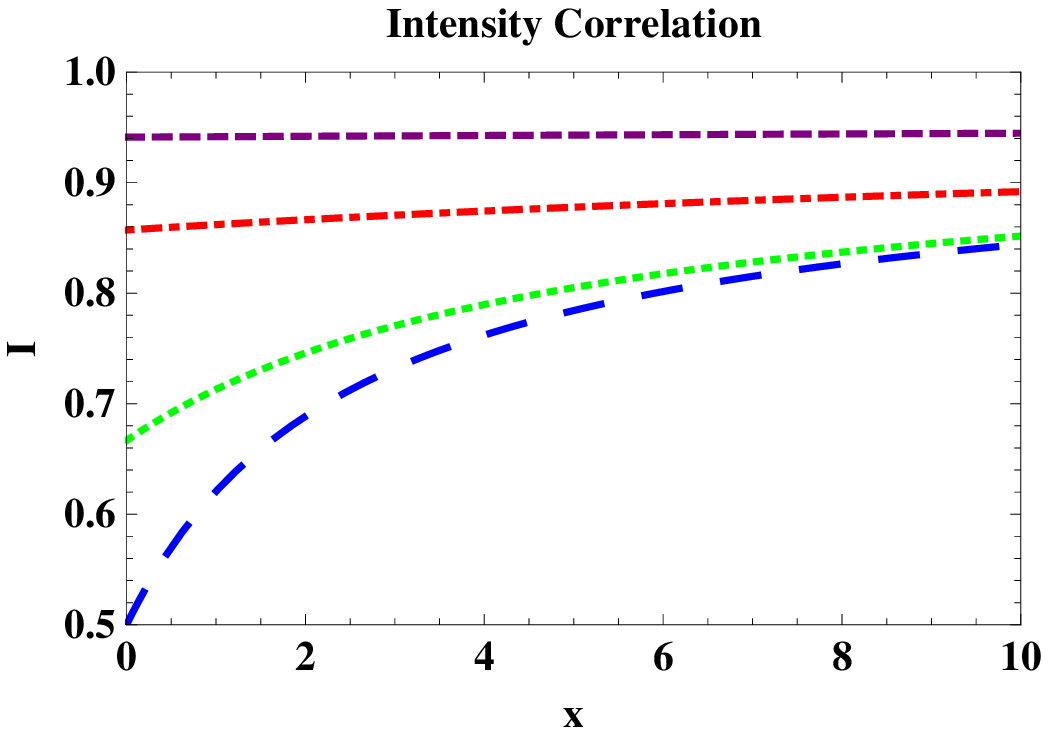}
\caption{$su(1,1)$ BGCS} \label{su11BGintcorr}
\end{minipage}
&
\begin{minipage}{3in}
\centering
\includegraphics[height=2in,width=2.8in]{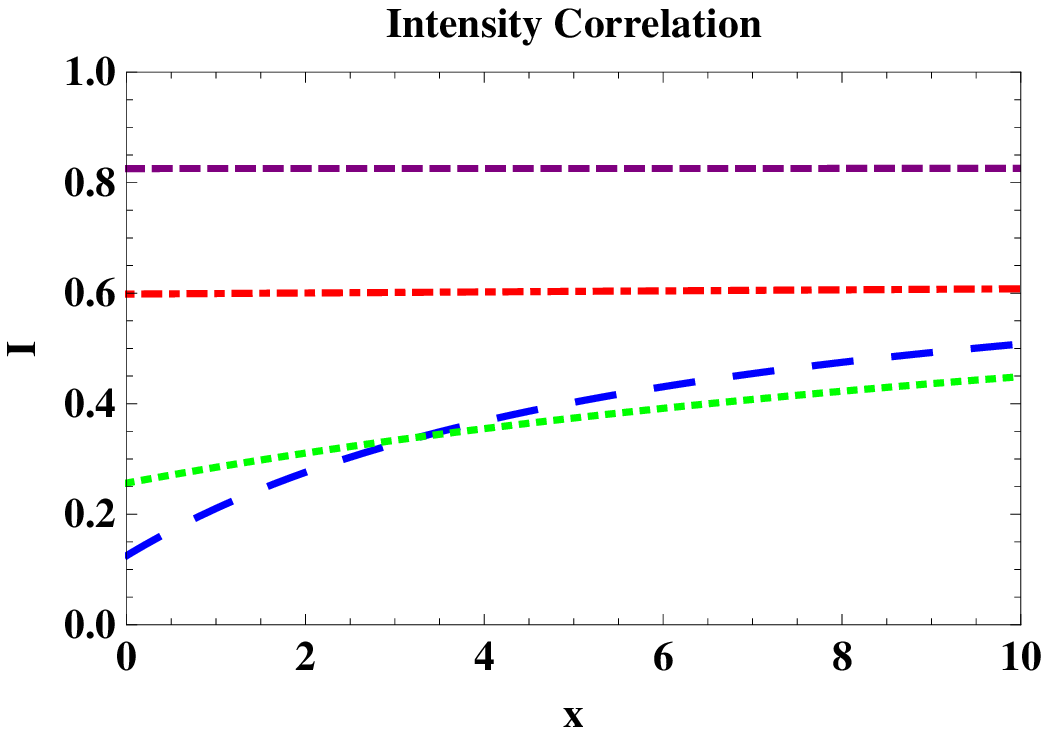}
\caption{Nonlinear $su(1,1)$ BGCS} \label{Nsu11BGintcorr}
\end{minipage}
\end{tabular}
\ec
\ef
\bf
\bc
\begin{tabular}{cc}
\begin{minipage}{3in}
\centering
\includegraphics[height=2in,width=2.8in]{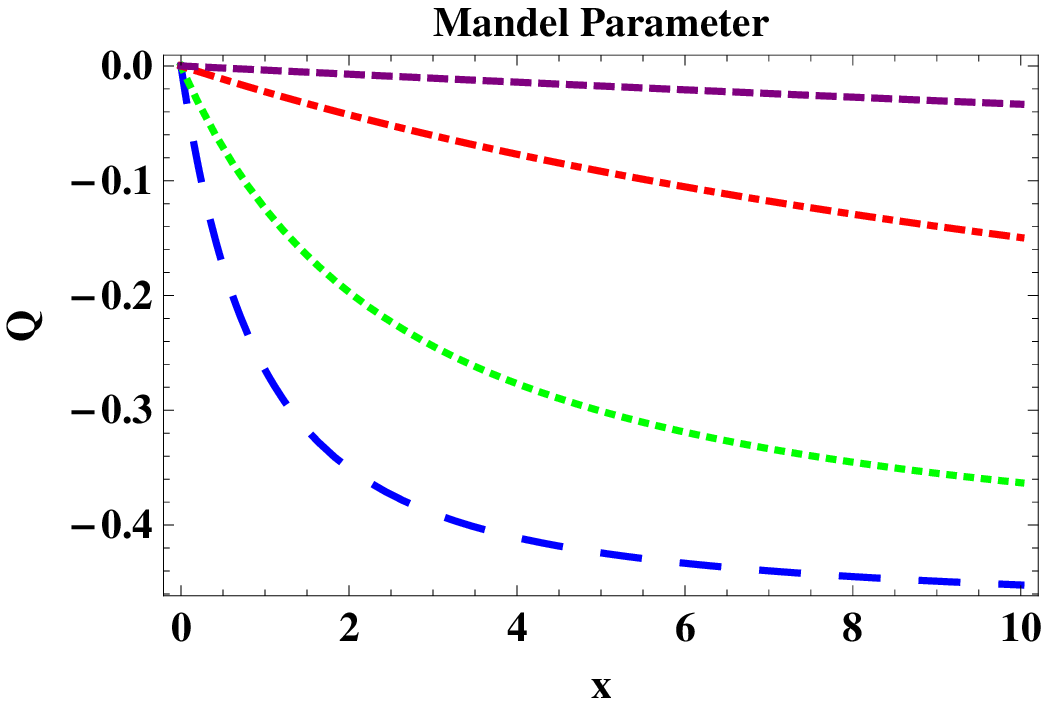}
\caption{$su(1,1)$ BGCS} \label{su11BGmandel}
\end{minipage}
&
\begin{minipage}{3in}
\centering
\includegraphics[height=2in,width=2.8in]{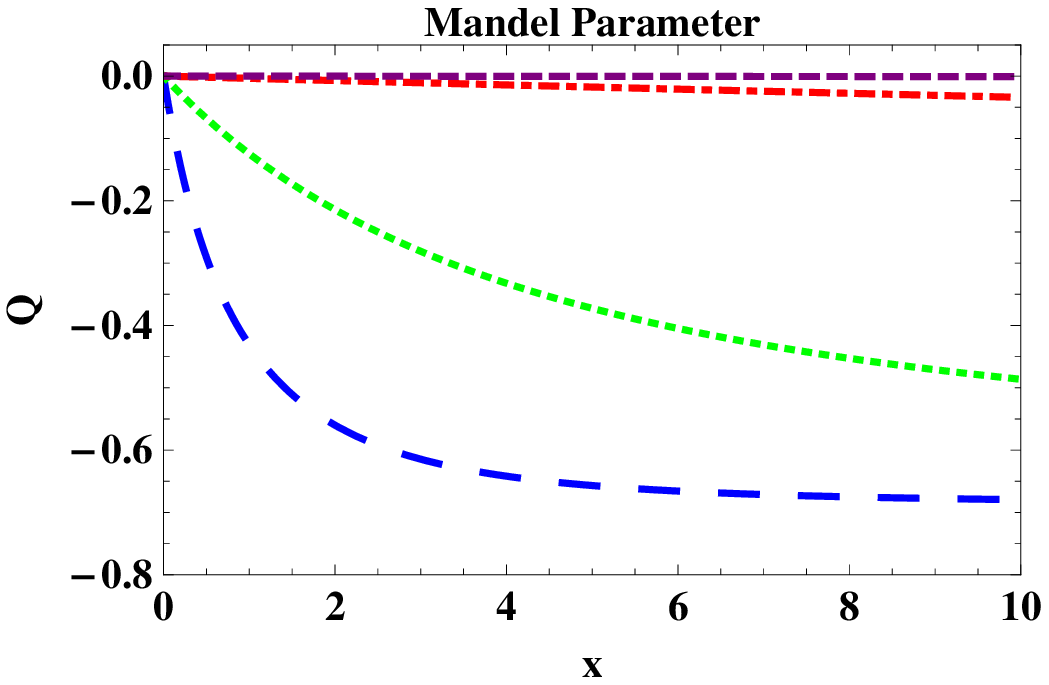}
\caption{Nonlinear $su(1,1)$ BGCS} \label{Nsu11BGmandel}
\end{minipage}
\end{tabular}
\ec
\ef

The graphs of figures (\ref{su11BGphotdist}), (\ref{su11BGmphotnumb}), (\ref{su11BGintcorr}), and (\ref{su11BGmandel}) are obtained from equations (\ref{photondist}), (\ref{mphotonumb}), (\ref{intcorr}) and (\ref{mandel}) respectively. These plots have been obtained using the  normalisation constant of $su(1,1)$ BGCS which is $N_1(x) = {_0}F_1[-;2k;y]/\Gamma(2k)$. These results were first presented in \cite{Brif-Stat1}, hence we will not discuss them any further in this work. We have reproduced them here for the sake of comparison with the nonlinear results. Figures (\ref{Nsu11BGphotdist}), (\ref{Nsu11BGmphotnumb}), (\ref{Nsu11BGintcorr}), and (\ref{Nsu11BGmandel}) are the graphs for the same physical quantities mentioned before but for the nonlinear $su(1,1)$ BGCS. From the intensity correlation and Mandel parameter plots, Figs. (\ref{Nsu11BGintcorr}) and (\ref{Nsu11BGmandel}) respectively, we can see that the statistics is sub-Poisonian hence we can call them sub-CS. Note from Fig. (\ref{Nsu11BGmandel}) how the result approaches the classical limit, $\mathcal{Q} < 0$ as $k$ increases unlike the $su(2)$ result, Fig. (\ref{Nsu2mandel}). Similarly the mean photon number plot Fig. (\ref{Nsu11BGmphotnumb}) behaves contrary to Fig. (\ref{Nsu2mphotnumb}). But the intensity correlation values approach $1$ in both the nonlinear $su(2)$ as well as $su(1,1)$ as $j,k$ increases, and this is depicted in Figs. (\ref{Nsu2intcorr})  and (\ref{Nsu11BGintcorr}) respectively. The photon number distribution plots indicate is shown in Fig. (\ref{Nsu11BGphotdist}).

\subsection{Deformed $su(1,1)$ PCS}

The photon number distribution, mean photon number, and the Mandel parameter are
\be
\mathcal{P}(z) =  \frac{N_p^{-1}(z) \ (2k)_n }{n! \prod_{i=1}^{2p-2} (1-b_i)_n} \ z^n
\ee
\be
\mathcal{N}(z) = z \ \frac{_1F_{2p-2}[2k+1; 2-b_1, 2-b_2, \cdots, 2-b_{2p-2};z]}{_1F_{2p-2}[2k; 1-b_1, 1-b_2, \cdots, 1-b_{2p-2};z]} \ \frac{(2k)}{\prod_{i=1}^{2p-2} (1-b_i)} \ .
\ee
Intensity correlation
\bea \no
\mathcal{I}(z) =   \frac{(2k+1)}{(2k)} \ \frac{\prod_{i=1}^{2p-2} (1-b_i)}{\prod_{i=1}^{2p-2} (2-b_i)} \ _1F_{2p-2}[2k; 1-b_1, 1-b_2, \cdots, 1-b_{2p-2};z] \\ \frac{_1F_{2p-2}[2k+2; 3-b_1, 3-b_2, \cdots, 3-b_{2p-2};z]}{_1F_{2p-2}[2k+1; 2-b_1, 2-b_2, \cdots, 2-b_{2p-2};z]^2} \ ,
\eea
and the Mandel parameter are
\bea \no 
\mathcal{Q}(z) = z \ \left[\frac{_1F_{2p-2}[2k+2; 3-b_1, 3-b_2, \cdots, 3-b_{2p-2};z]}{_1F_{2p-2}[2k+1; 2-b_1, 2-b_2, \cdots, 2-b_{2p-2};z]} \ \frac{(2k+1) }{\prod_{i=1}^{2p-2} (2-b_i)} \right. \\ \left. - \ \frac{_1F_{2p-2}[2k+1; 2-b_1, 2-b_2, \cdots, 2-b_{2p-2};z]}{_1F_{2p-1}[2k; 1-b_1, 1-b_2, \cdots, 1-b_{2p-2};z]} \ \frac{(2k)}{\prod_{i=1}^{2p-2} (1-b_i)} \right] \ .
\eea
\bf
\bc
\begin{tabular}{cc}
\begin{minipage}{3in}
\centering
\includegraphics[height=2in,width=2.8in]{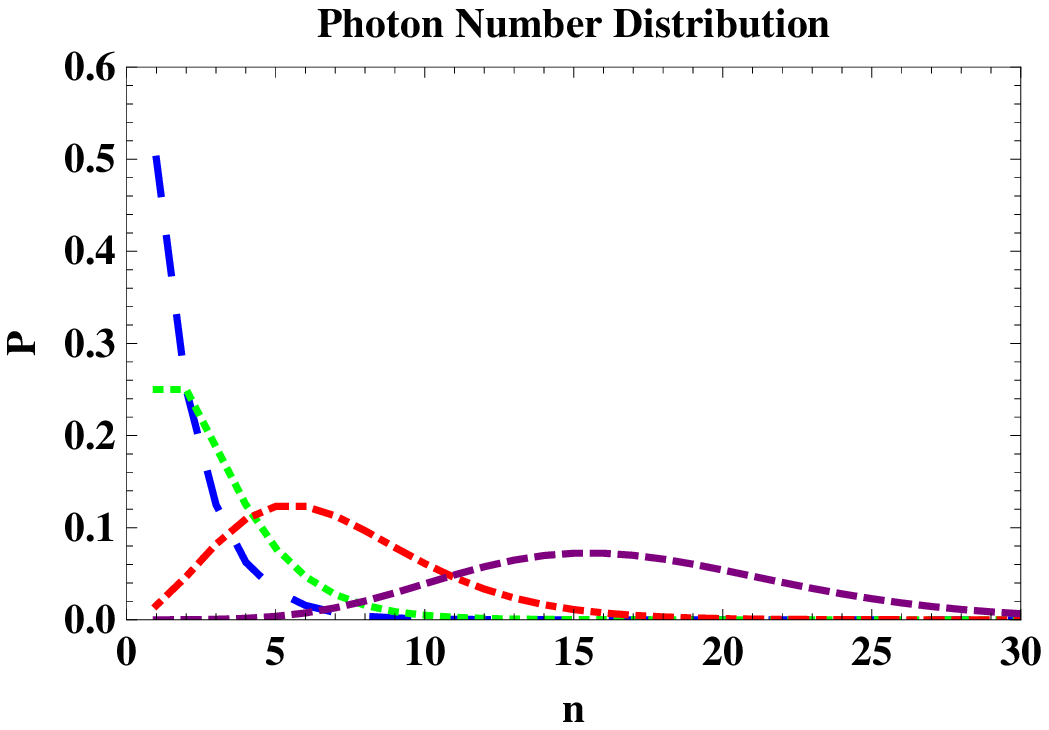}
\caption{$su(1,1)$ PCS} \label{su11PCphotdist}
\end{minipage}
&
\begin{minipage}{3in}
\centering
\includegraphics[height=2in,width=2.8in]{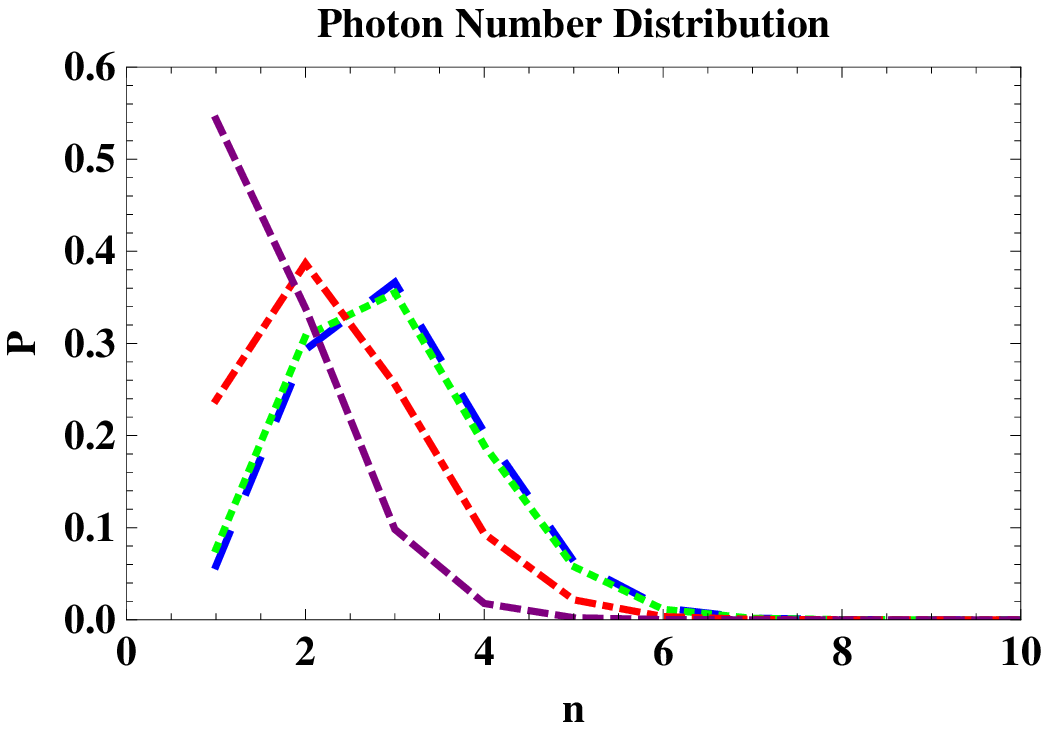}
\caption{Nonlinear $su(1,1)$ PCS} \label{Nsu11PCphotdist}
\end{minipage}
\end{tabular}
\ec
\ef
\bf
\bc
\begin{tabular}{cc}
\begin{minipage}{3in}
\centering
\includegraphics[height=2in,width=2.8in]{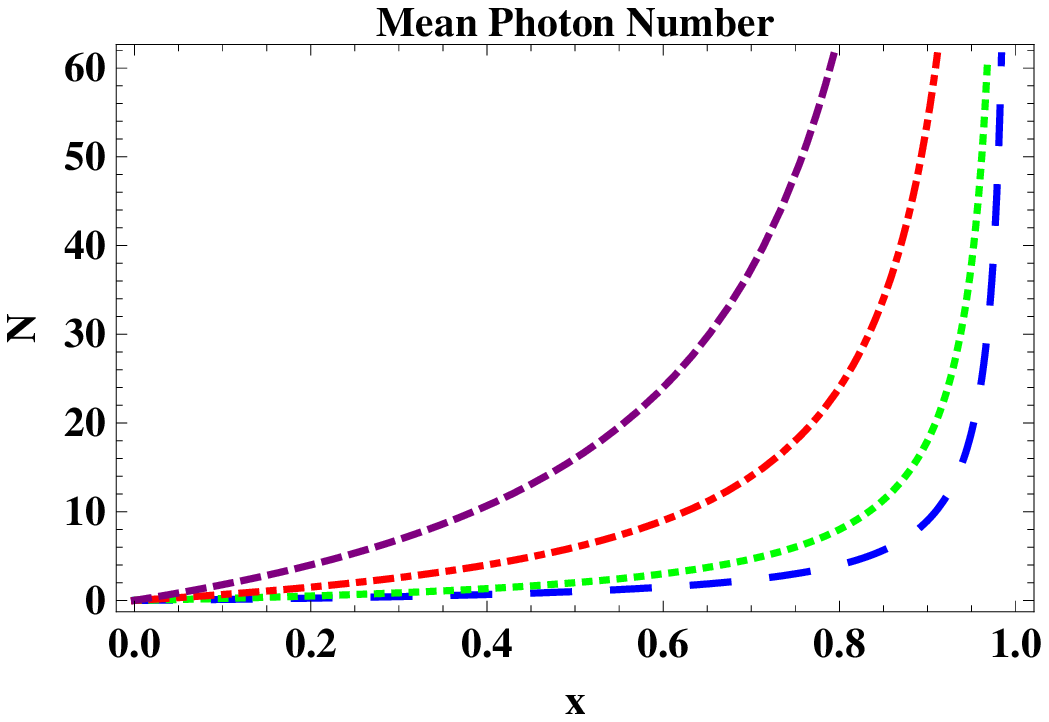}
\caption{$su(1,1)$ PCS} \label{su11PCmphotnumb}
\end{minipage}
&
\begin{minipage}{3in}
\centering
\includegraphics[height=2in,width=2.8in]{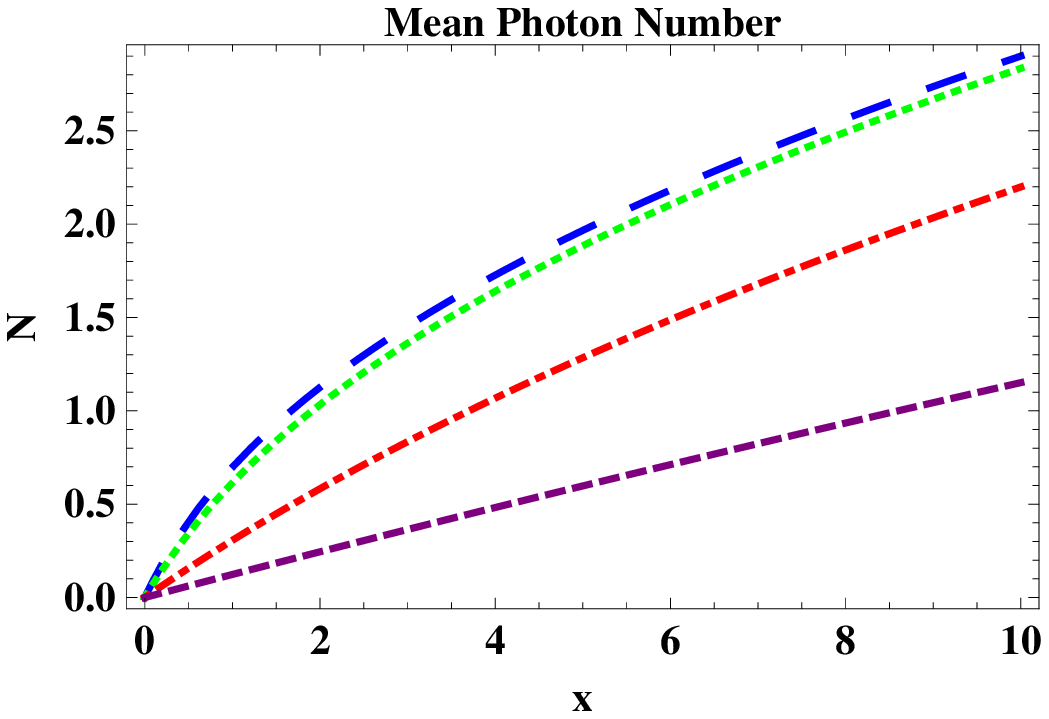}
\caption{Nonlinear $su(1,1)$ PCS} \label{Nsu11PCmphotnumb}
\end{minipage}
\end{tabular}
\ec
\ef
\bf
\bc
\begin{tabular}{cc}
\begin{minipage}{3in}
\centering
\includegraphics[height=2in,width=2.8in]{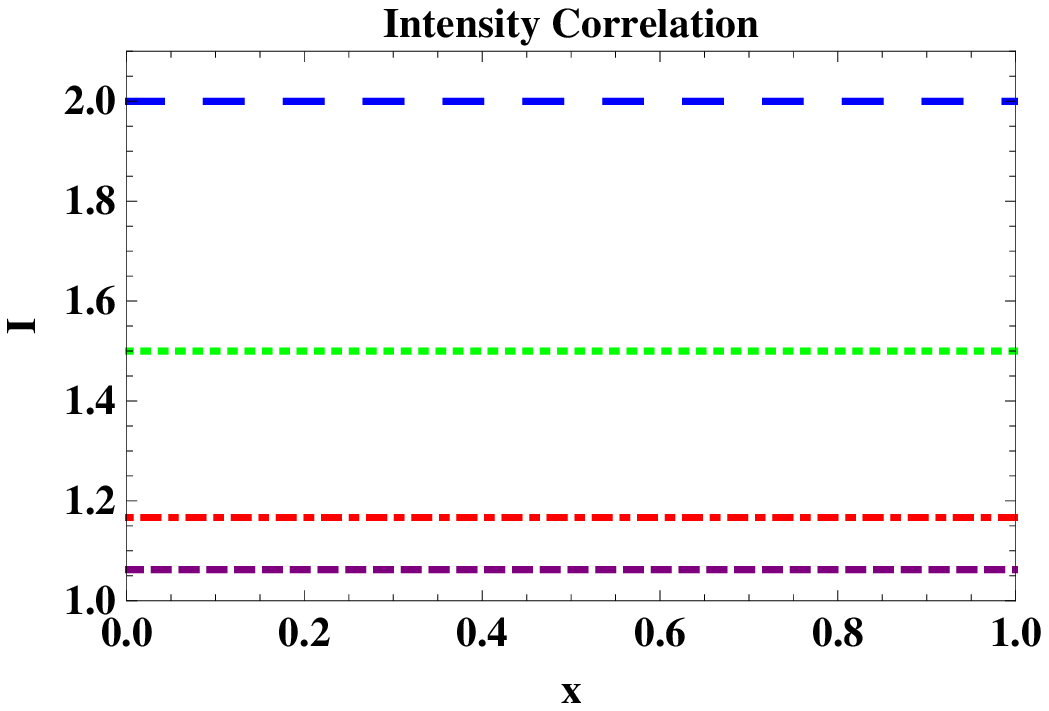}
\caption{$su(1,1)$ PCS} \label{su11PCintcorr}
\end{minipage}
&
\begin{minipage}{3in}
\centering
\includegraphics[height=2in,width=2.8in]{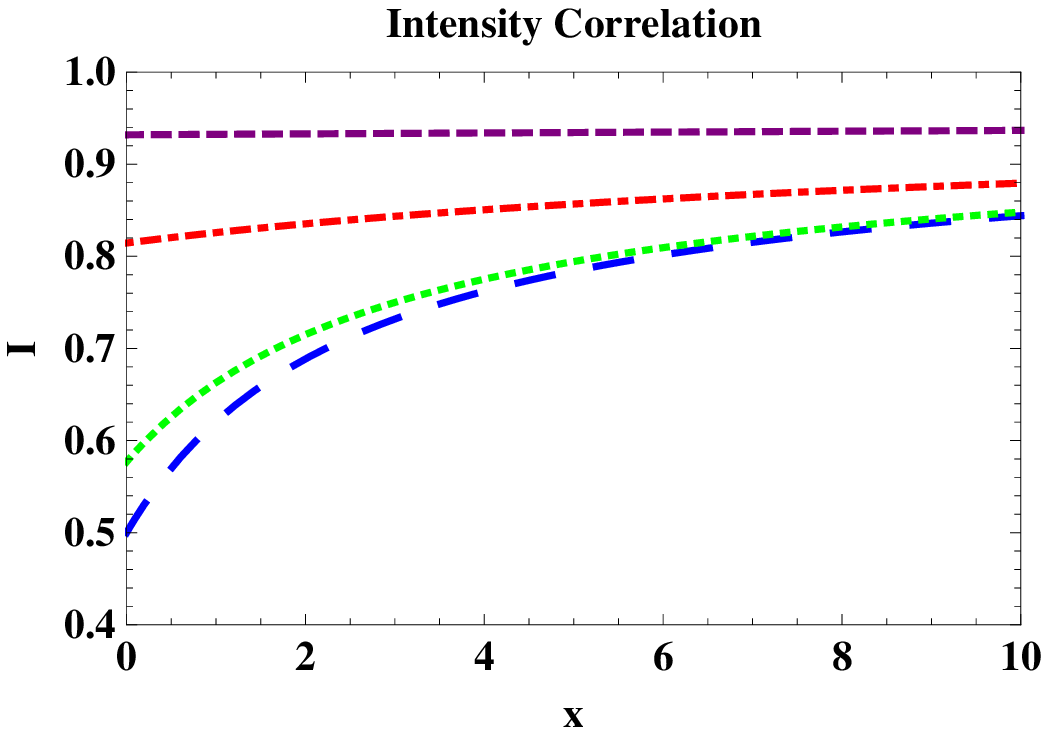}
\caption{Nonlinear $su(1,1)$ PCS} \label{Nsu11PCintcorr}
\end{minipage}
\end{tabular}
\ec
\ef
\bf
\bc
\begin{tabular}{cc}
\begin{minipage}{3in}
\centering
\includegraphics[height=2in,width=2.8in]{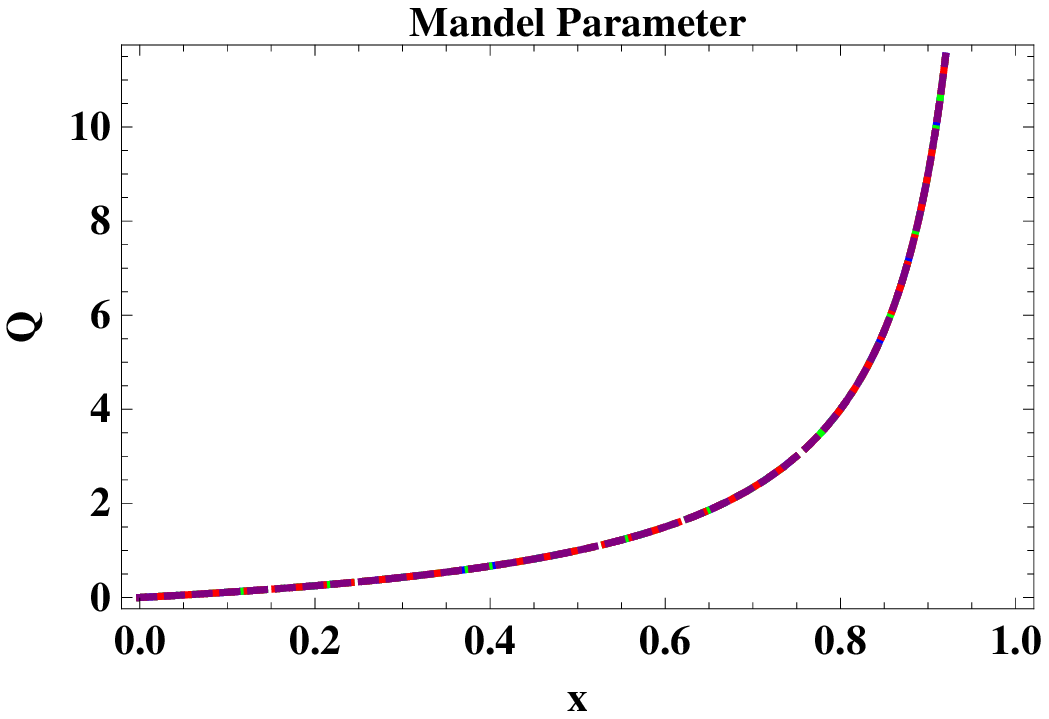}
\caption{$su(1,1)$ PCS} \label{su11PCmandel}
\end{minipage}
&
\begin{minipage}{3in}
\centering
\includegraphics[height=2in,width=2.8in]{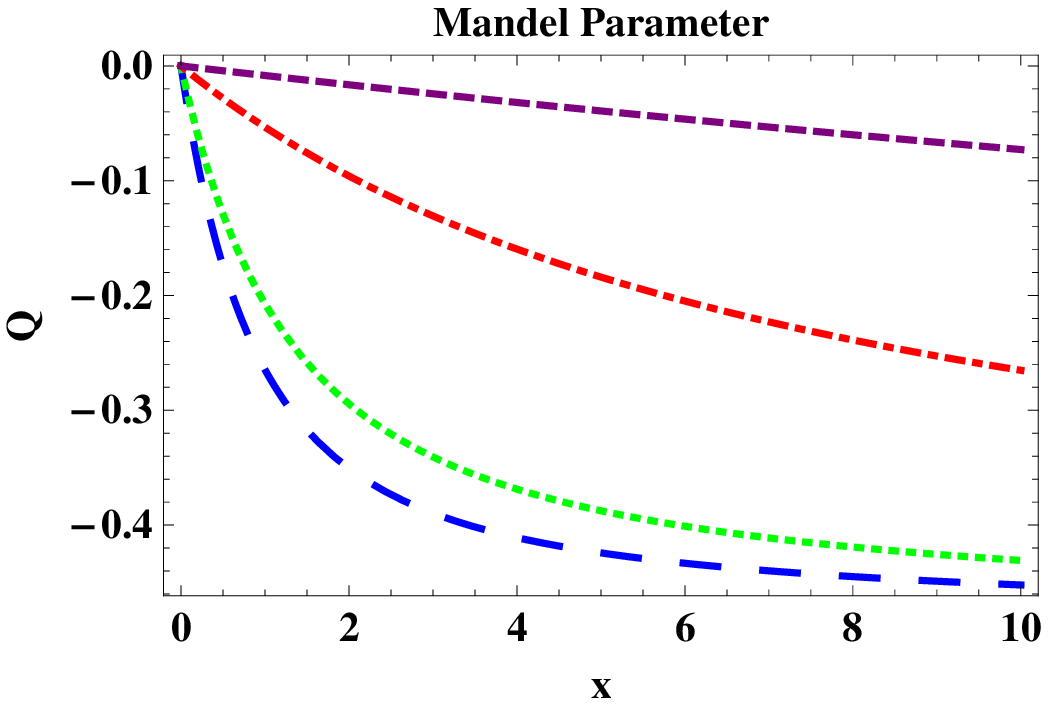}
\caption{Nonlinear $su(1,1)$ PCS} \label{Nsu11PCmandel}
\end{minipage}
\end{tabular}
\ec
\ef

The graphs shown in figures (\ref{su11PCphotdist}), (\ref{su11PCmphotnumb}), (\ref{su11PCintcorr}), and (\ref{su11PCmandel}) are obtained from equations (\ref{photondist}), (\ref{mphotonumb}), (\ref{intcorr}) and (\ref{mandel}) respectively. These plots have been obtained using the  normalisation constant of $su(1,1)$ PCS which is $N_1(z) = {_1}F_0[2k;-;z]$. The $su(1,1)$ CS is super-Poissonian since $\mathcal{I} > 1$ and $\mathcal{Q} > 0$, Figs. (\ref{su11PCintcorr}) and (\ref{su11PCmandel}),  whereas for the non-linear case, Figs. (\ref{Nsu11PCintcorr}) and (\ref{Nsu11PCmandel}), it is sub-Poissonian for reasons already mentioned. The photon distribution is provided in Figs. (\ref{su11PCphotdist}) and Fig. (\ref{Nsu11PCphotdist}). Also note how the behaviour of the intensity correlation for linear and the nonlinear case is contrary for different values of $k$. Figures (\ref{su11PCmphotnumb}) and (\ref{Nsu11PCmphotnumb}) depict the mean photon number. The photon number distribution for $su(1,1)$, Fig. (\ref{su11PCphotdist}) is a negative binomial distribution \cite{GSA,Vourdas}.

\section{Nonlinear metrics}

In the previous section we have discussed the statistical properties of the CS. In the present, we discuss some geometric aspects of the same. That displacement operator CS is in one-to-one correspondence with the elements of the coset space is well-known from the work of Perelomov. Thus the geometrical properties of the coset manifold are inherited by the CS. But this runs into difficulty when one considers the BGCS or those CS that do not possess a group theoretical construction, like the one being considered in the present work. One can then still have some qualitative understanding of the geometry of such states by calculating the so called metric factor. For a discussion and derivation of this consult Ref \cite{Klauder}. But definitely a more detailed analysis along the lines of \cite{Lane} needs to be done and we hope to return to it in future.

The metric factor can be expressed entirely in terms of the normalisation constant of the CS:
\be \label{metric}
\omega(\bar{x}) = \frac{N^\prime(\bar{x})}{N(\bar{x})} + \bar{x} \ \left(\frac{N^{\prime \prime}(\bar{x})}{N(\bar{x})}-\frac{N^{\prime 2}(\bar{x})}{N^2(\bar{x})} \right).
\ee
Using the normalisation of equation (\ref{normnlinsu2}) in Eq. (\ref{metric}) gives
\bea \no
\omega(x) & = & - \ (-2j) \ \frac{_{2p-1}F_{0}[-2j+1, 2-a_1, 2-a_2, \cdots, 2-a_{2p-2};-;-x]}{_{2p-1}F_{0}[-2j,1-a_1,1-a_2, \cdots, 1-a_{2p-2};-;-x]} \ \prod_{i=1}^{2p-2} (1-a_i) \\ \no
& + & \ x \ (-2j)_2 \ \frac{_{2p-1}F_{0}[-2j+2, 3-a_1, 3-a_2, \cdots, 3-a_{2p-2};-;-x]}{_{2p-1}F_{0}[-2j, 1-a_1, 1-a_2, \cdots, 1-a_{2p-2};-;-x]} \ \prod_{i=1}^{2p-2} (1-a_i)_2 \\
& - & \ x \ (-2j)^2 \ \frac{_{2p-1}F_{0}[-2j+1,2-a_1,2-a_2, \cdots, 2 - a_{2p-2};-;-x]^2}{_{2p-1}F_{0}[-2j,1-a_1,1-a_2, \cdots, 1-a_{2p-2};-;-x]^2} \ \Big[\prod_{i=1}^{2p-2} (1-a_i) \Big]^2 \ .
\eea
the metric factor associated with the BGCS normalisation constant in Eq. (\ref{nlinbgcsnorm}) reads
\bea \no
\omega(y) & = & \frac{1}{(2k)} \ \frac{_0F_{2p-1}[-; 2k+1, 2-b_1, 2-b_2, \cdots, 2-b_{2p-2};y]}{_0F_{2p-1}[-; 2k, 1-b_1, 1-b_2, \cdots, 1-b_{2p-2};y]} \ \frac{1}{\prod_{i=1}^{2p-2} (1-b_i)} \\ \no
& + & \frac{y}{(2k)_2} \ \frac{_0F_{2p-1}[-; 2k+2, 3-b_1, 3-b_2, \cdots, 3-b_{2p-2};y]}{_0F_{2p-1}[-; 2k, 1-b_1, 1-b_2, \cdots, 1-b_{2p-2};y]} \ \frac{1}{\prod_{i=1}^{2p-2} (1-b_i)_2} \\
& - & \frac{y}{(2k)^2} \ \frac{_0F_{2p-1}[-; 2k+1, 2-b_1, 2-b_2, \cdots, 2-b_{2p-2};y]^2}{_0F_{2p-1}[-; 2k, 1-b_1, 1-b_2, \cdots, 1-b_{2p-2};y]^2} \ \frac{1}{\big[\prod_{i=1}^{2p-2} (1-b_i)\big]^2} \ .
\eea
A similar calculation for the normalisation constant of Eq. (\ref{nlinpcsnorm}) leads to
\bea \no
\omega(z) & = & \frac{_1F_{2p-2}[2k+1; 2-b_1, 2-b_2, \cdots, 2-b_{2p-2};z]}{_1F_{2p-2}[2k; 1-b_1, 1-b_2, \cdots, 1-b_{2p-2};z]} \ \frac{(2k)}{\prod_{i=1}^{2p-2} (1-b_i)} \\ \no
& + & z \ \frac{_1F_{2p-2}[2k+2; 3-b_1, 3-b_2, \cdots, 3-b_{2p-2};z]}{_1F_{2p-2}[2k; 1-b_1, 1-b_2, \cdots, 1-b_{2p-2};z]} \ \frac{(2k)_2}{\prod_{i=1}^{2p-2} (1-b_i)_2} \\
& - & z \ \frac{_1F_{2p-2}[2k+1; 2-b_1, 2-b_2, \cdots, 2-b_{2p-2};z]^2}{_1F_{2p-2}[2k; 1-b_1, 1-b_2, \cdots, 1-b_{2p-2};z]^2} \ \frac{(2k)^2}{\big[\prod_{i=1}^{2p-2} (1-b_i)\big]^2} \ .
\eea
\bf
\bc
\begin{tabular}{cc}
\begin{minipage}{3in}
\centering
\includegraphics[height=2in,width=2.8in]{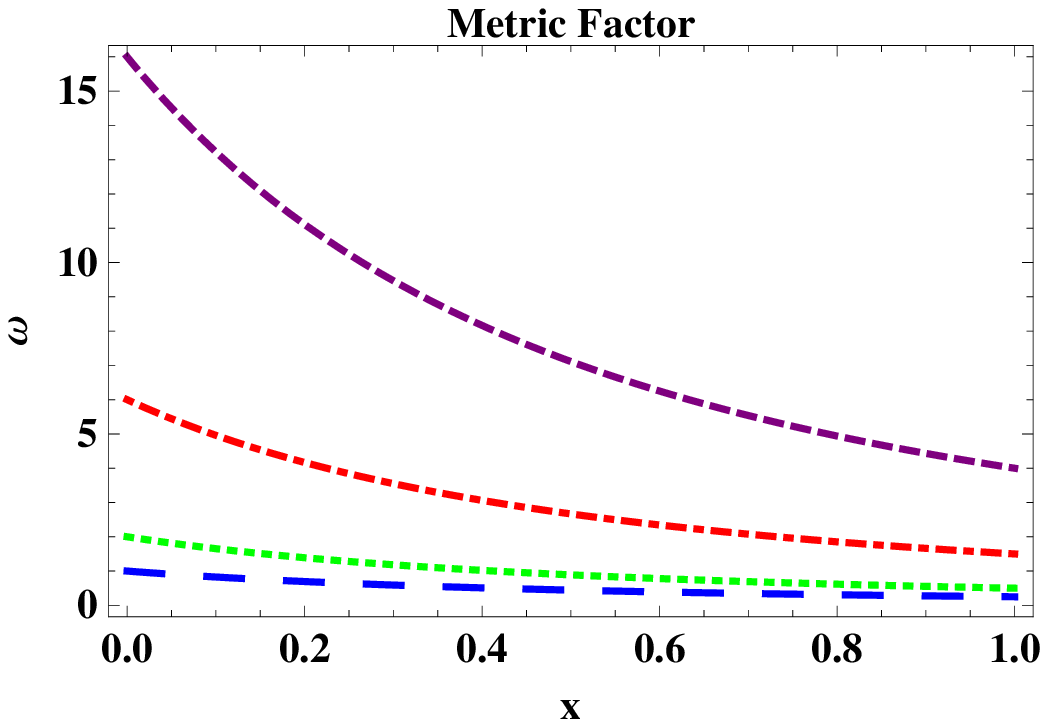}
\caption{$su(2)$ PCS} \label{su2metric}
\end{minipage}
&
\begin{minipage}{3in}
\centering
\includegraphics[height=2in,width=2.8in]{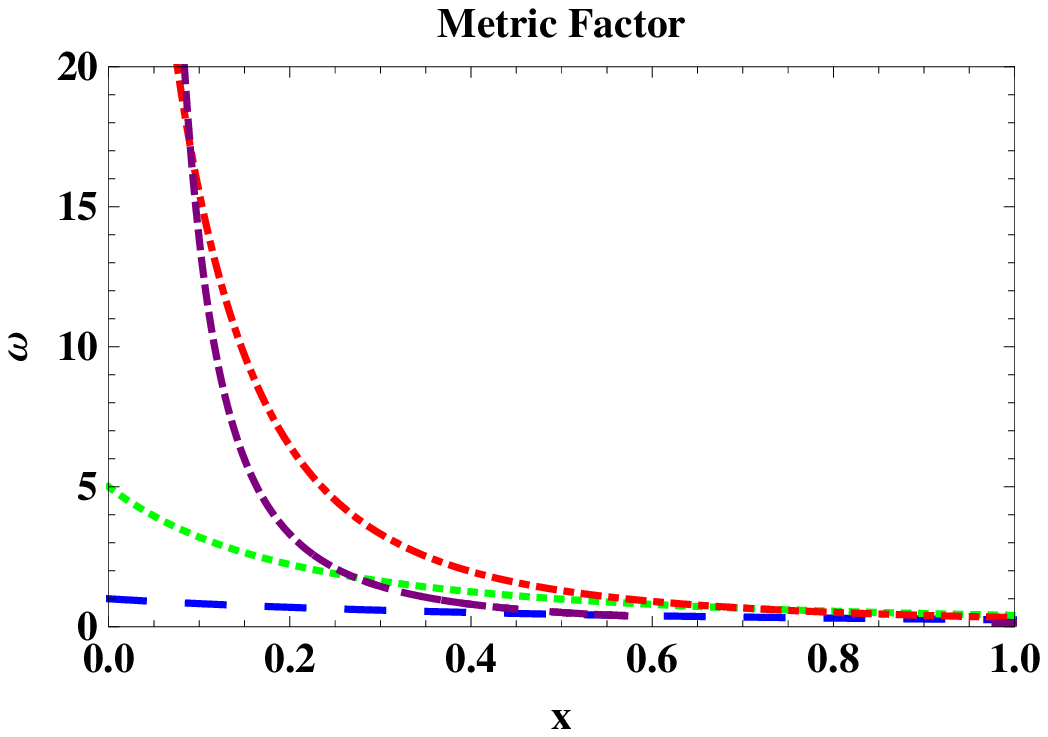}
\caption{Nonlinear $su(2)$ PCS} \label{Nsu2metric}
\end{minipage}
\end{tabular}
\ec
\ef
\bf
\bc
\begin{tabular}{cc}
\begin{minipage}{3in}
\centering
\includegraphics[height=2in,width=2.8in]{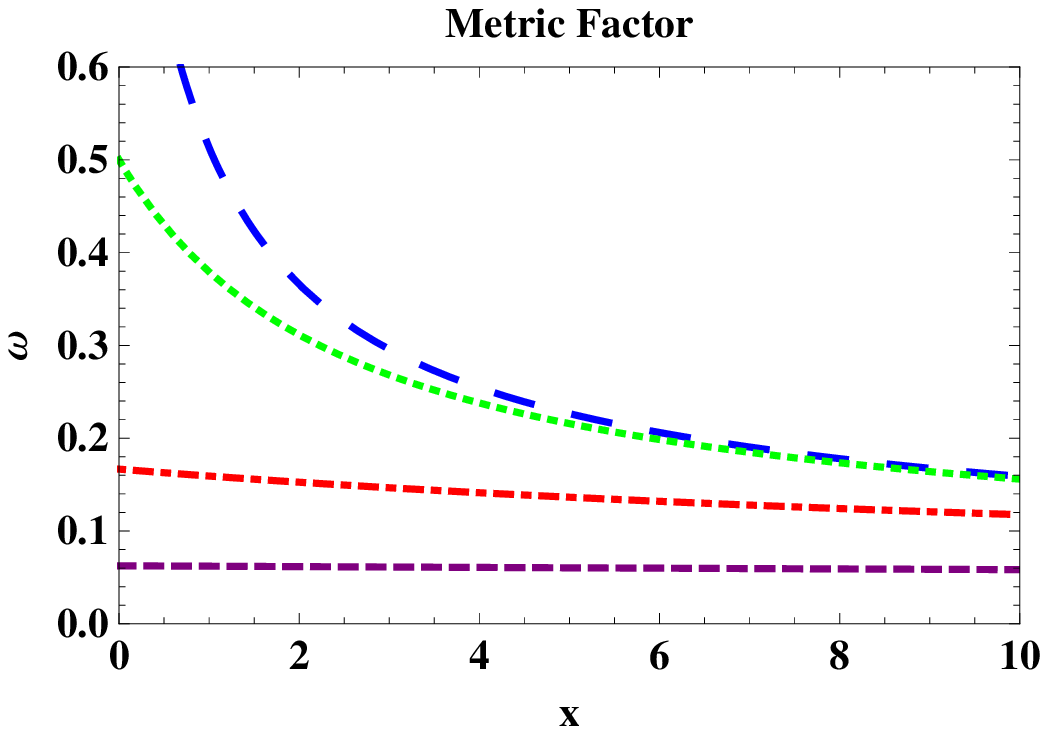}
\caption{$su(1,1)$ BGCS} \label{su11BGmetric}
\end{minipage}
&
\begin{minipage}{3in}
\centering
\includegraphics[height=2in,width=2.8in]{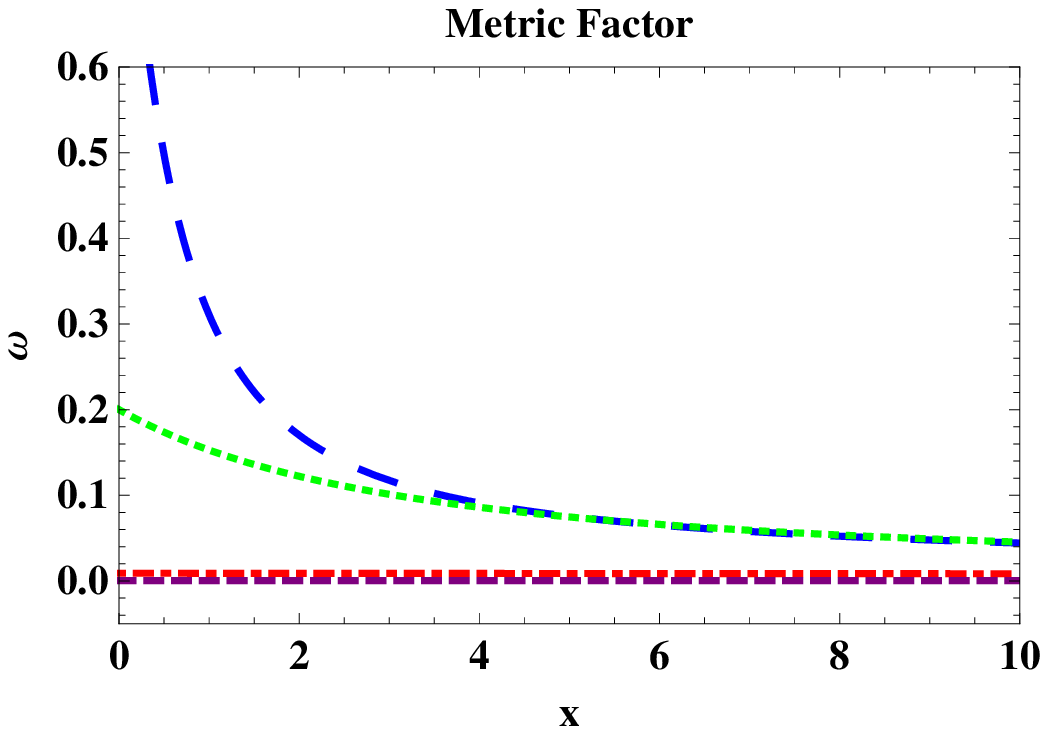}
\caption{Nonlinear $su(1,1)$ BGCS} \label{Nsu11BGmetric}
\end{minipage}
\end{tabular}
\ec
\ef
\bf
\bc
\begin{tabular}{cc}
\begin{minipage}{3in}
\centering
\includegraphics[height=2in,width=2.8in]{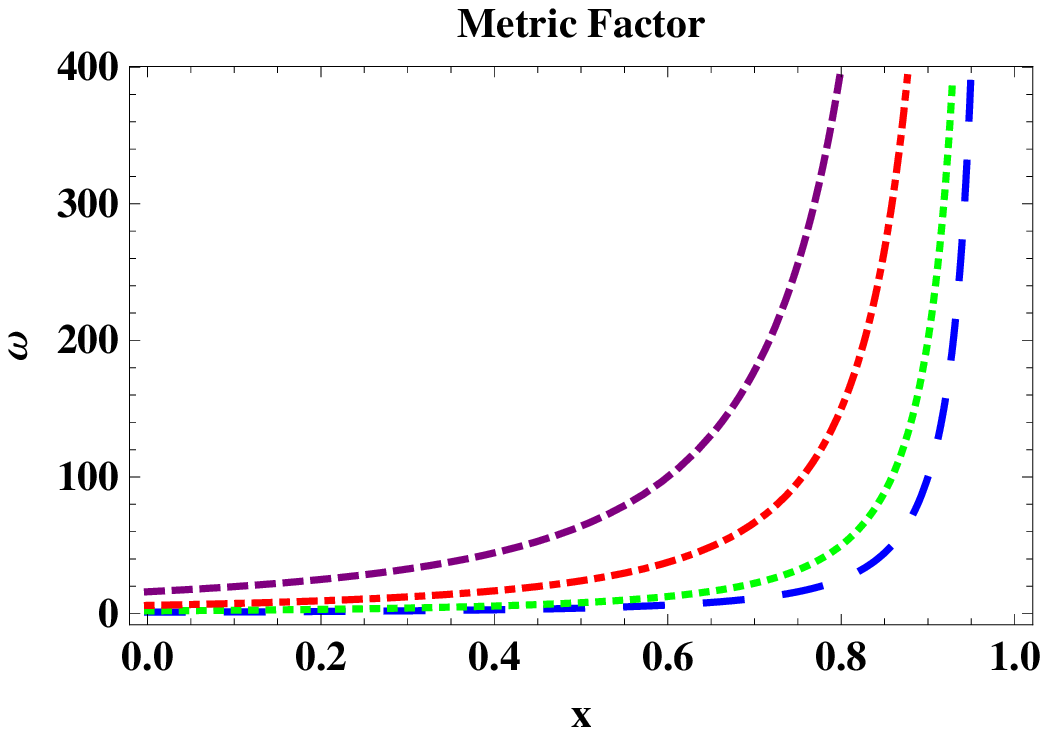}
\caption{$su(1,1)$ PCS} \label{su11PCmetric}
\end{minipage}
&
\begin{minipage}{3in}
\centering
\includegraphics[height=2in,width=2.8in]{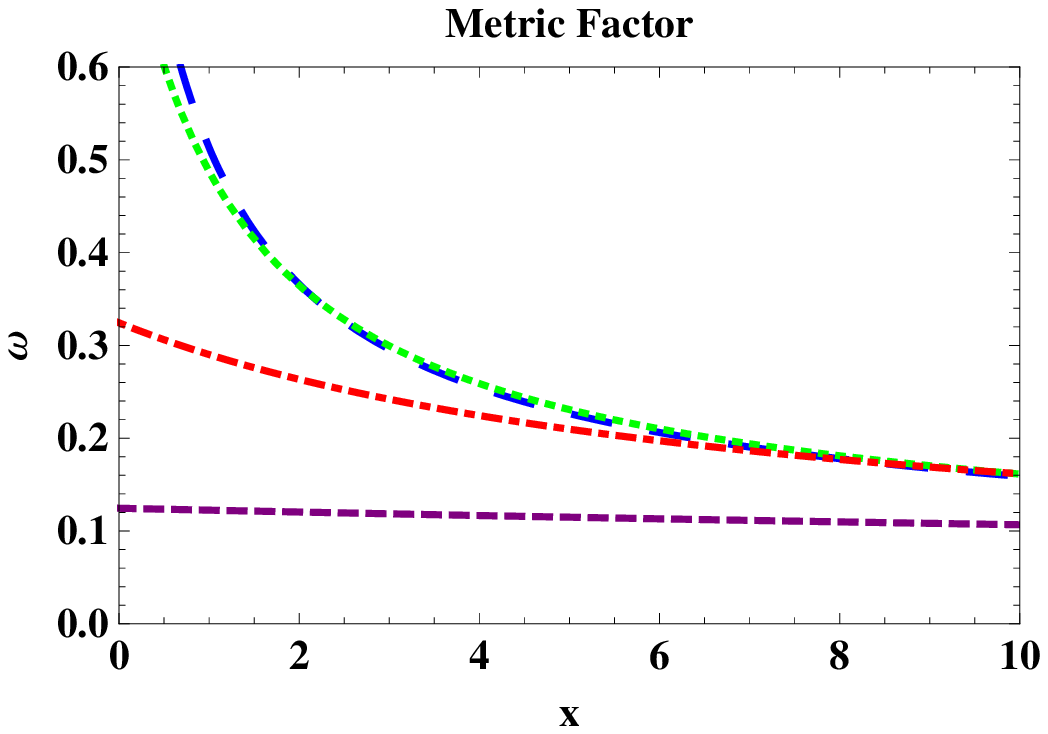}
\caption{Nonlinear $su(1,1)$ PCS} \label{Nsu11PCmetric}
\end{minipage}
\end{tabular}
\ec
\ef

We have plotted the metric factors of $su(2)$ and nonlinear $su(2)$ in Figs. (\ref{su2metric}) and (\ref{Nsu2metric}) respectively. Similarly Figs. (\ref{su11BGmetric}) and (\ref{Nsu11BGmetric}) are the curves depicting the metric factors of $su(1,1)$ BGCS and nonlinear BGCS. Metric factors for $su(1,1)$ PCS and nonlinear PCS are depicted in Figs. (\ref{su11PCmetric}) and (\ref{Nsu11PCmetric}). All metrics except the $su(1,1)$ PCS become asymptotically flat. That is as $x$ increases the matric factor becomes zero. In fact for the oscillator CS it is zero. A non zero metric factor indicates a curvature. In the case of Figs. (\ref{su2metric}) and (\ref{Nsu2metric}) for small values of $j$ the metric is nearly flat which is not the case for large $j$ and small $x$. In fact for the nonlinear case the curvature is very sharp for small $x$ and large values of $j$. In the BGCS plots these conclusions are reversed. Namely from Figs. (\ref{su11BGmetric}) and (\ref{Nsu11BGmetric}) one can notice that the metric is nearly flat for large values of $k$. For nonlinear $su(1,1)$, Fig. (\ref{su11BGmetric}), the metric factor is identically zero for $k=8$. The behaviour of nonlinear $su(1,1)$ PCS, Fig. (\ref{Nsu11PCmetric}), is similar to the $su(1,1)$ BGCSv at least for large $k$. The $su(1,1)$ PCS behaves as expected when $x=1$ since the group manifold is a hyperboloid.

\section{The Berry's Phase}

Berry \cite{Berry} has shown that under a cyclic adiabatic evolution the wave function of a Hamiltonian, that depends on slowly varying external parameters, picks up an additional phase apart form the dynamical one. This indeed was a surprising result since, the extra phase depends only on the geometry of the cycle. This result was subsequently generalised to scenarios such as non-adiabatic \cite{Aharonov}, noncyclic and even non-unitary \cite{Samuel} evolutions. It was also extended to include the non-Abelian case \cite{Wilczek}. Finally a surprising result showed that the phase need not be be acquired via dynamical evolution but the origin can be a kinematic one \cite{Mukunda}.

Consider a Hamiltonian $H(R)$ with slowly varying parameters $R$ and a discrete spectrum:
\be
 H(R)\ \vert n, R\rangle = E_n \ \vert n, R \rangle,
\ee
where $\vert n, R \rangle$ are the normalisable eigenstates. Under a adiabatic, cyclic, and time periodic evolution of the parameter $R$, the initial state $\vert n, R\rangle$ acquires an extra phase and is given as \cite{Berry}
\be
 \gamma_n = i \int_{0}^{T} dt \ \langle n, R \vert \frac{d}{dt} \vert n, R \rangle .
\ee
The above expression may alternatively be expressed in the form
\be \label{bee}
 \gamma_n = i \oint_{0}^{T} dt \ \langle n, R \vert \nabla_R \vert n, R \rangle .
\ee
In the rest of the section we will calculate the geometric phase for CS that are eigenstates of some Hamiltonian. We will not be interested in the particular form of the Hamiltonian. But, the essential fact that will be made use of is that the spectrum be discrete and the states be labeled by a non-negative integer $n$. The investigation of the Berry's phase for CS was first initiated in \cite{Chaturvedi} and later extended to include the squeezed states in \cite{Lakshmi}.

\subsection{Deformed su(2) algebra}

We start our discussion of Berry-phase with the following identification $\vert n, R\rangle \equiv \vert j, \zeta\rangle$ where $\vert j, \zeta\rangle$ is as given in Eq. (\ref{pcssu2}). After some algebra we get
\be \label{b1}
\langle \zeta, j\vert\frac{d}{dt}\vert j,\zeta\rangle  = -\frac{\alpha_p}{2} \ N_p^{-1}(x) \ \frac{dN_p(x)}{dx} \ \left(\frac{d\zeta}{dt}\zeta^* + \zeta\frac{d\zeta^*}{dt}\right)
+ \frac{d\zeta}{dt} \ \langle \zeta, j\vert J_+\vert j,\zeta \rangle \ .
\ee
All that remains to be done is to calculate the derivative of the normalisation constant and the matrix element of $J_+$ . The derivative is easy to find, therefore we only give details on calculation of the matrix element
\begin{eqnarray} \no
\langle \zeta, j\vert J_+\vert j,\zeta \rangle & = & \zeta^*N_p^{-1}(x) \sum_{n=0}^{2j-1}\frac{\vert\zeta\vert^{2n}}{n!} \frac{\left[\psi_{n+1}\right]!}{(n+1)!} \\ \no
& = & 2 j\zeta^* N_p^{-1}(x) \sum_{n=0}^{2j-1}\frac{\vert\zeta\vert^{2n}}{n!} \frac{(2j-1)!}{(2j-1-n)!}[\chi_{n+1}]! \\
& = & 2j\alpha_p \zeta^* N_p^{-1}(x) \sum_{n=0}^{2j-1} x^{n}  \left(\begin{array}{c}  2j-1 \\ n \end{array}\right) \prod_{i=1}^{2p-2} (1-a_i)_{n+1} \ .
\end{eqnarray}
In arriving at the above expression we have made use of the fact that
\be
(2j-n)! = \frac{2j!}{(2j-1-n)!} = (n+1)!\left(\begin{array}{c} 2j \\ n+1 \end{array}\right)
\ee
Now to proceed further we use the recurrence relation pertaining to the Pochhammer symbol $(O)_{n+1} = O(O + 1)_n$, thus we get
\be
\langle \zeta, j\vert J_+ \vert j,\zeta \rangle  =  2 j \zeta^* N_p^{-1}(x) \prod_{i=1}^{2p-2} \alpha_p (1-a_i) \sum_{n=0}^{2j-1} \frac{(-x)^{n}}{n!} (-2j+1)_n \prod_{i=1}^{2p-2} (2-a_i)_{n} \ .
\ee
We have used the following
\be
\left(\begin{array}{c} 2j-1 \\ n \end{array}\right) = \frac{(-1)^{n}}{n!} (-2j+1)_n \ .
 \ee
The sum can be easily recognized to be the HS. Thus the final answer for the matrix element turns out to be
\be
\langle \zeta, j\vert J_+ \vert j,\zeta \rangle =  2 j \zeta^* \ [\chi_{1}]! \ \frac{  {_{2p-1}F_{0}}[-2j+1, 2-a_1, 2-a_2, 2-a_{2p-2};-;-x]}
{ {_{2p-1}F_{0}}[-2j, 1-a_1, 1-a_2, \cdots, 1-a_{2p-2};-;-x] }
\ee
Using the above expression and the derivative of the normalisation constant, in Eq. (\ref{b1}) we arrive at
\be
\langle \zeta, j\vert \frac{d}{dt}\vert j,\zeta \rangle = \frac{(2j) \ [\chi_{1}]!}{2} \left(\zeta^*\frac{d\zeta}{dt} - \frac{d\zeta^*}{dt} \zeta \right) 
\frac{{_{2p-1}F_{0}}[-2j+1, 2-a_1, 2-a_2, \cdots, 2-a_{2p-2};-;-x]}{ {_{2p-1}F_{0}}[-2j, 1-a_1, 1-a_2, \cdots, 1-a_{2p-2};-;-x]} \ .
\ee
The final result for the Berry's phase can be obtained by substituting the above equation in Eq. (\ref{bee}). It can be seen easily that when the nonlinearity does not exist, it reproduces the result of the $su(2)$ derived in \cite{Chaturvedi}.

\subsection{Deformed su(1,1) BGCS}

One can calculate without much effort the result for the BGCS, Eq. (\ref{bgcs})  and it is
\bea \no 
\langle \xi, k \vert \frac{d}{dt}\vert k,\xi \rangle = -\frac{N_p^{-1}(y)}{2\beta_p} \frac{dN_p(y)}{dy} \left(\frac{d\xi^\ast}{dt} \xi + \xi^\ast \frac{d\xi}{dt}\right)
+ \frac{\xi^*}{(2k)[\rho_{1}]!} \frac{d\xi}{dt} \\  \label{b3} \frac{{_{0}F_{2p-1}}[-; 2k+1, 2-b_1, 2-b_2, \cdots, 2-b_{2p-2};y]}{{_{0}F_{2p-1}}[-; 2k, 1-b_1, 1-b_2, \cdots, 1-b_{2p-2};y]} \ .
\eea
The above can be cast in the final form
\be
\langle \xi, k\vert \frac{d}{dt}\vert k,\xi \rangle = \frac{1}{2(2k)[\rho_{1}]!} \left(\xi^*\frac{d\xi}{dt} - \frac{d\xi^*}{dt} \xi \right)
\frac{{_{0}F_{2p-1}}[-; 2k+1, 2-b_1, 2-b_2, \cdots, 2-b_{2p-2};y]}{ {_{0}F_{2p-1}}[-; 2k, 1-b_1, 1-b_2, \cdots, 1-b_{2p-2};y]} \ .
\ee
Once again Eq. (\ref{bee}) in conjunction with the above result gives the Berry's phase.

\subsection{Deformed su(1,1) PCS}

Finally we give the Berry phase calculation for the nonlinear $su(1,1)$ PCS
\be
\langle \eta, k \vert \frac{d}{dt} \vert k,\eta \rangle = -\frac{N_p^{-1}(z)}{2\beta_p} \frac{dN_p(z)}{dz} \left(\frac{d\eta^\ast}{dt} \eta + \eta^\ast \frac{d\eta}{dt}\right) + \frac{N_p^{-1}(z)}{\eta} \frac{d\eta}{dt} \sum_{n=0}^\infty \frac{n (2k)_n}{n! [\rho_n]!} \ \eta^n \ .
\ee
The summation can be performed and the final answer takes the form
\begin{eqnarray}
\langle \eta, k\vert \frac{d}{dt}\vert k,\eta \rangle & = & \frac{(2k)}{2[\rho_{1}]!} \left(\eta^*\frac{d\eta}{dt} - \frac{d\eta^*}{dt} \eta \right) \nonumber \\
&& \frac{  {_1F_{2p-2}}[2k+1; 2-b_1, 2-b_2, \cdots, 2-b_{2p-2};-;z]}{ {_1F_{2p-2}}[2k; 1-b_1, 1-b_2, \cdots, 1-b_{2p-2};-;z]} \ .
 \end{eqnarray}

\section{Discussion and Conclusion}

In the present work we have studied various aspects of CS constructed for compact as well as as noncompact nonlinear algebras. Specifically, we have calculated the photon number distribution, mean photon number, intensity correlation, Mandel parameter and the metric factor. We have also shown that the nonlinear $su(1,1)$ PCS and the BGCS are related via Laplace transform. It has been clearly shown from the plots that the $su(2)$, nonlinear $su(2)$ PCS, as well as $su(1,1)$, nonlinear $su(1,1)$ BGCS, and nonlinear $su(1,1)$ PCS have sub-Poissonian statistics. Only the $su(1,1)$ PCS has a super-Poissonian statistics. We have also studied the metric factor associated with the linear as well as the nonlinear CS. But an in depth analytical study is needed. In the present work have just given a qualitative feel for the geometrical structure via the metric factor. The Berry's phase corresponding to all the three different CS has been obtained. It will be nice to give a geometrical interpretation for the phases. The primary difficulty one faces in the study of metrics and the geometrical underpinning of Berry's phase, is the absence of a group structure for the nonlinear algebras. This in turn translates into the problem of existence of a proper exponential map.

An interesting spin-off of the present study is the following. The nonlinear CS that have been used in the present work can be viewed as special cases of CS $\vert p,q,\alpha \rangle$ whose normalisation constant is given by $_p F_q$. These CS were constructed in \cite{Appl}, motivated by \cite{Klauder}, as arising from the Stieltjes and
Hausdorff moment problems. Now, one can try to see if these states have any underlying nonlinear algebraic structure. Furthermore one can study the geometrical structure of the CS constructed in \cite{Appl}. Also the Berry phase can be calculated for the same.

As future work one can repeat all the calculations presented in this work for other non-linear algebras. The non-linear algebra that we have in mind is the quadratic one. This might be easy since the CS have already been constructed. Another interesting algebra that one can consider is the Delbeq-Quesne (DQ) algebra \cite{Quesne1}.

Let $X_+,~X_-,~ X_0$ be the generators. The DQ algebra is then defined by the following commutation relations: 
\be 
[X_+, X_-] = P(X_0) \ , \quad [X_0, X_+] = G(X_0) \, X_+ \ , \quad [X_0, X_-] = - X_- \ G(X_0). 
\ee 
Here $P$ and $G$ are polynomials of the diagonal operator, $X_0$. To the best of our knowledge CS for this algebra has not been constructed. Thus for this algebra one can first construct the CS. Once that is done we can then do a similar study as that presented in this work.

We hope to to return to some of the problems mentioned above at a later date.

\end{document}